\documentclass[prl,aps,twocolumn,amsfonts,showpacs,superscriptaddress]{revtex4-1}

\pdfoutput=1

\usepackage{amsmath,amssymb}
\usepackage{epsfig}
\usepackage{graphicx}
\usepackage{hyperref}

\usepackage{comment}

\begin{document}

\title{Stationary holographic plasma quenches and numerical methods for non-Killing horizons}

\author{Pau Figueras}
\affiliation{DAMTP, Centre for Mathematical Sciences, University of Cambridge, Wilberforce Road, Cambridge CB3 0WA, U.K.}

\author{Toby Wiseman}
\affiliation{Theoretical Physics Group, Blackett Laboratory, Imperial College, London SW7 2AZ, U.K.}

\date{December 2012}

\begin{abstract}
We explore use of the harmonic Einstein equations to numerically find stationary black holes where the problem is posed on an ingoing slice that extends into the interior of the black hole.  Requiring no boundary conditions at the horizon beyond smoothness of the metric, this method may be applied for horizons that are not Killing. 
As a non-trivial illustration we find black holes which, via AdS-CFT, describe a
time-independent CFT plasma flowing through a static spacetime which asymptotes to Minkowski in the flow's past and future, with a varying spatial geometry in-between.
These are the first explicit examples of stationary black holes which do not have Killing horizons. 
When the CFT spacetime slowly varies, the CFT stress tensor derived from gravity is well described by viscous hydrodynamics. For fast variation it is not, 
and the solutions are stationary analogs of dynamical quenches,
with the plasma being suddenly driven out of equilibrium.
We find evidence these flows become unstable for sufficiently strong quenches, and speculate the instability may be turbulent.
\end{abstract}


\maketitle

\section{Introduction}

Due to the remarkable Anti de Sitter-Conformal Field Theory (AdS-CFT) correspondence 
\cite{Maldacena:1997re,Gubser:GTC,Witten:AdSholo}, the behaviour of black holes in asymptotically AdS spacetimes is equivalent 
to the behaviour of hot plasma in certain strongly coupled CFTs. 
Since these black holes may have planar horizons, 
they admit perturbations of arbitrarily long wavelength which give rise to the hydrodynamic behaviour expected of the CFT plasma 
\cite{Kovtun:2003wp,Buchel:2003tz,Kovtun:2004de,Baier:2007ix,Bhattacharyya:2008jc}.
Perturbations on short scales correspond to microscopic plasma behaviour beyond hydrodynamics. 
As this currently cannot be computed directly in strongly coupled CFTs, gravity provides an entirely new computational tool,
 \cite{Hartnoll:Lectures,McGreevy:Notes}.
This has been exploited for dynamical quenches where the CFT is abruptly perturbed \cite{Danielsson:1999zt,Giddings:1999zu,Bhattacharyya:2009uu,Das:2010yw,
  Albash:2010mv} and where the dual spacetime is determined by numerical dynamical gravity \cite{Chesler:2008hg,Murata:Noneq,Bizon:2011gg,Garfinkle:2011hm,Bantilan:2012vu,Buchel:Thermal,Bhaseen:2012gg}. 
 
 Here we study an analog of a dynamical quench, where the CFT state is time independent.
 We consider stationary black holes dual to a time independent
relativistic plasma flow through a static spacetime. This asymptotes to Minkowski in the flow's past and future, but in-between the spatial geometry varies in the flow direction. 
The flow, initially in equilibrium, is forced out of equilibrium in response to passing through the curved spacetime region, before returning to equilibrium afterwards.
For slowly varying spacetimes (with respect to the length scale set by the local temperature) such flows are well described by hydrodynamics \cite{Bhattacharyya:2008ji}. For quick variation they probe behaviour beyond hydrodynamics, 
and are stationary analogs to dynamical quenches. 

These black holes are of a qualitatively new variety, being the first explicit examples of stationary black holes  that do not have Killing horizons, and hence do not move rigidly 
\footnote{It is interesting to contrast this with the solutions of \cite{Dias:2011at} which have a Killing horizon, but are neither stationary nor axisymmetric if one considers the metric \emph{and} matter. However, considering the metric alone, the solution is stationary and does rigidly rotate. We refer to rigidity with reference to the metric alone in this work.}. 
The rigidity theorem states 
that if a stationary horizon is compact, it is also Killing \cite{Hawking:1973uf,Hollands:2006rj,Moncrief:2008mr}. Our black holes have non-compact horizons and evade this theorem, and since the dual plasma flows in a direction which is not a symmetry, these horizons are not Killing. 
Other stationary non-Killing horizons have been conjectured in the AdS-CFT context; `flowing funnels' \cite{Hubeny:2009ru,Fischetti:2012ps} and `plasma shocks' \cite{Khlebnikov:2010yt,Khlebnikov:2011ka}.  So far only related solutions with Killing horizons have been found \cite{Figueras:2011va,Santos:2012he}.

\section{Harmonic Einstein equations}

Consider a Lorentzian stationary solution to the Einstein equations where the stationary Killing vector field $T$ is globally timelike. We consider the purely gravitational  case $R_{\mu\nu} = \Lambda g_{\mu\nu}$, although generalisation to include matter is straightforward. 
We adapt coordinates, $x^\mu = ( t, x^i)$, so $T = \partial / \partial t$ and the metric $g_{\mu\nu}$ is,
\begin{eqnarray}
\label{eq:stat}
ds^2 = -N(x) (dt + A_i(x) dx^i)^2 + b_{ij}(x) dx^i dx^j \, .
\end{eqnarray}
The spacetime is Lorentzian so $\det g_{\mu\nu} = - N \det b_{ij} <0$, and $T$ is globally timelike so $N(x)>0$ and thus $b_{ij}(x)$ is a Riemannian metric. 
In order to obtain a well posed problem we must eliminate the coordinate invariance.
Instead of solving the Einstein equation, we solve the `harmonic' or `DeTurck' Einstein equations \cite{Headrick:2009pv,Adam:2011dn,Wiseman:2011by},
\begin{eqnarray}
\label{eq:harmonic}
R^H_{\mu\nu} \equiv   R_{\mu\nu} - \nabla_{(\mu} \xi_{\nu)} = \Lambda g_{\mu\nu}
\end{eqnarray}
where,
$
\xi^\mu = g^{\alpha\beta} \left( \Gamma^\mu_{~~\alpha\beta} - \bar{\Gamma}^\mu_{~~\alpha\beta} \right)
$
is constructed from a fixed reference connection $\bar{\Gamma}^\mu_{~~\alpha\beta}$ on the manifold, which here we take to be the connection of a  reference metric $\bar{g}_{\mu\nu}$. The two derivative  part of these equations is governed by the operator $b^{ij} \partial_i \partial_j$,
and since $b_{ij}$ is a Riemannian metric the harmonic Einstein equation is elliptic.
 
For suitable boundary conditions
the whole system may be solved as a standard elliptic boundary value problem. We want solutions with $\xi^\mu = 0$, which is a coordinate gauge condition analogous to generalised harmonic gauge in dynamical numerical GR \cite{Garfinkle:2001ni}, and must ensure our boundary conditions are compatible with this.
In certain cases one may prove $\xi^\mu$ must vanish \cite{Figueras:2011va}. In general, `soliton' solutions with $\xi^\mu \ne 0$ may exist. However for an elliptic problem solutions are locally unique, and hence one may easily distinguish whether a solution found has vanishing $\xi^\mu$ or not.
The system may be solved by relaxation
which is related to Ricci flow. Alternatively after discretization it can be solved by the Newton method given an initial guess.

\section{Old method for Killing horizons}

For a stationary black hole $T = \partial/\partial t$ is no longer globally timelike, being spacelike inside the horizon, or outside if an ergoregion exists. Hence $b_{ij}$ must become Lorentzian, and the problem inside the horizon and ergoregion  naively appears hyperbolic. 

A previous method \cite{Adam:2011dn} focussed on retaining ellipticity by assuming a Killing horizon that rigidly moves. 
We assume Killing vectors $R_{a} = \partial / \partial y^a$ exist which commute with themselves and $T$. For constants $\Omega^a$ we take  $K = T + \Omega^a R_a$ to generate the Killing horizon and  rigid motion of the spacetime. The metric can then be written as,
\begin{eqnarray}
ds^2 &= &G_{AB}(x) (dy^A + A^A_a(x) dx^a) (dy^B + A^B_b(x) dx^b) \nonumber\\
&\,&+ b_{ab}(x) dx^a dx^b
\end{eqnarray}
with $y^A = (t, y^a)$. Now $G_{AB}$ is Lorentzian outside the horizon (even in an ergoregion), and degenerates at the horizon or axes of symmetry of the $R_a$. Correspondingly $b_{ab}(x)$ can be chosen to be Riemannian on, and in the exterior of, the horizon.
These coordinates yield a slice of the spacetime that intersects the bifurcation surface of the Killing horizon.
Since the principle part of $R^H$ is governed by $b^{ab} \partial_a \partial_b$ the p.d.e. system is elliptic posed on the base geometry with coordinates $x^a$ - the `orbit space'. The boundaries of this base are where $G_{AB}$ degenerates and smoothness determines boundary conditions there \cite{Hollands:2007aj,Harmark:2009dh,Adam:2011dn}. The surface gravity $\kappa$ and (angular) velocities of the horizon $\Omega^a$ are prescribed in these boundary conditions. For regularity the reference metric must also have a Killing horizon at the same location with the same $\kappa$ and $\Omega^a$. Hence we may think of the reference metric as specifying these moduli of the solution.

\section{New method for non-Killing horizons}

Suppose we are interested in stationary black holes that do not have a Killing horizon. 
For a non-Killing horizon we cannot assume existence of a bifurcation surface and regular past horizon. In the new approach we now describe we no longer require the problem to be elliptic. We take the general stationary ansatz \eqref{eq:stat} and pose the harmonic Einstein equations on an ingoing slice (analogous to that of Eddington-Finklestein) that intersects the future horizon and extends into the black hole interior. For the metric \eqref{eq:stat} in ingoing coordinates $g_{\mu\nu}$ is regular at the future horizon, so $\det g_{\mu\nu} = -N \det b_{ij} < 0$ on the future horizon and its exterior. Thus $b_{ij}$ is elliptic in the exterior of the horizon or ergoregion (ie. where $N>0$) and is hyperbolic inside these (where $N<0$). Such a problem is analogous to mixed hyperbolic elliptic p.d.e.s in fluid dynamics. Whilst the problem will have hyperbolic character inside the horizon and ergoregion we may still solve it using the Newton method as before. Interestingly the Ricci flow method appears also to work, but we will not explore that here.
All components of $\xi^\mu = 0$ give non-trivial gauge conditions; the $\xi^i$ are associated to coordinate freedom in $x^i$ and $\xi^t$ to the freedom $t \to t + f(x^i)$.

An important difference to the old method is that since the problem is hyperbolic in the interior of the horizon, at the innermost points of our domain we impose only the harmonic Einstein equations and no boundary condition. The requirement that the metric is smooth in our domain is sufficient to ensure regularity of the horizon in ingoing coordinates.  
Starting from a smooth initial guess near a solution, then the Ricci flow and Newton method will preserve smoothness, since both update the metric using the harmonic Ricci tensor which will also be smooth.
We implicitly assume the physically reasonable statement that asymptotic boundary conditions together with future horizon regularity define a locally unique stationary black hole solution, 
up to moduli of the solution (such as mass). 
This is true for Killing horizons as can be seen from the elliptic nature of the p.d.e.s discussed earlier. Indeed the  black hole uniqueness theorems show in many cases global uniqueness.
For stationary non-Killing horizons we assume local uniqueness here, but emphasise we know of no proof. We note this is  the basis of the fluid/gravity correspondence \cite{Bhattacharyya:2008jc,Baier:2007ix}.
It is the horizon rather than the ergosurface where smoothness must be imposed, even though the ergosurface determines the transition of character of the p.d.e.s. This is analogous to a stationary scalar field in the Kerr background, where one explicitly sees the scalar equation has regular singular behaviour at the horizon, and hence it is smoothness there that constrains the solutions
\footnote{This is to be contrasted with the canonical mixed hyperbolic
elliptic equation, the Tricomi equation, $\left(\partial^2 / \partial y^2 + y \, \partial^2 / \partial x^2 \right) f = 0$ where smoothness at $y = 0$ (the analog of the horizon or ergosurface) imposes no condition on the solution.}.

A second key difference with the old method is 
that the reference metric, while selecting the coordinate system a solution is presented in, no longer specifies the surface gravity or velocities of the horizon. These moduli must be fixed by appropriate boundary conditions.

For a reader interested in implementing the method we provide a toy example in appendix A; finding Schwarzschild assuming static spherical symmetry. The old and  new methods are given and contrasted.

\section{Holographic plasma quenches}

An example application of the method above is to find stationary black holes that are locally asymptotically $AdS_4$, and are relevant in AdS-CFT to describe CFT stationary plasma flows in a non-trivial geometry. These are Einstein metrics solving $R_{\mu\nu} = - \frac{3}{l^2} g_{\mu\nu}$. We choose units so that the AdS length $l=1$. These geometries have a conformal boundary whose conformal class we are free to specify
and corresponds to the spacetime that the CFT is defined on. We choose a static metric,
\begin{eqnarray}
\label{eq:boundarymetric}
ds^2 = - dt^2 + d\rho^2 + \sigma(\rho) dy^2
\end{eqnarray}
where $\sigma(\rho)$ deforms the spatial geometry breaking the translation symmetry in $\rho$ as,
\begin{eqnarray}
\sigma(\rho) = \frac{1}{2} \left[ 1 + \frac{\alpha}{2} \Big( 1 + \tanh(\beta \rho) \Big) \right] 
\label{eq:sigma}
\end{eqnarray}
for constants $\alpha, \beta$, and the geometry asymptotes to Minkowski for $\rho \to \pm \infty$.
We take the CFT plasma to be stationary, homogeneous in $y$, and flowing from $\rho = - \infty$ to $+ \infty$.
We expect the plasma flow in the asymptotic Minkowski regions $\rho \to \pm \infty$ to become homogeneous, and correspondingly the dual to become a homogeneous black brane in these limits. However, since 
the spatial geometry of
the boundary metric depends non-trivially on the direction that the plasma flows in, the dual black hole must have velocity in a direction which
is not associated to
an isometry. Hence the plasma flow is inhomogeneous, and the dual black hole does not have a Killing horizon.

We take the ingoing plasma to be subsonic with velocity $v_0 < 1/\sqrt{2}$ and temperature $T_0$.
Using the holographic fluid/gravity correspondence \cite{Baier:2007ix,Bhattacharyya:2008jc} provided the boundary metric gradients are sufficiently small, meaning $\beta / T_0 \to 0$, then the plasma behaves as an ideal fluid. The first deviation from ideal behaviour is due to shear viscosity.  
Upon increasing $\beta / T_0$ towards unity, one expects the derivative expansion of hydrodynamics to break down completely in the region where the boundary metric is highly curved, as microscopic physics is required to describe small scale plasma phenomena. As we shall see, this microscopic breakdown of viscous hydrodynamics is indeed captured by the dual gravity black hole,
and these solutions represent stationary flowing plasma quenches.

We write an ansatz for these metrics as,
\begin{eqnarray}
\label{eq:metric}
ds^2 = \frac{1}{z^2} \left( - T dt^2 + 2 V dt dz + 2 U dt d\rho + A dz^2 \right. \nonumber \\
\left. + B \left( d\rho + F dz \right)^2 + S dy^2 \right)
\end{eqnarray}
with the functions $T,V,B,S,U,F,A$ being smooth (or at least $C^2$) in $\rho$ and $z$.
The locally $AdS_4$ boundary is at $z = 0$, and we impose the boundary conditions such that
\begin{eqnarray}
\label{eq:adsbc}
T = V = A = B = 1 \; , \, U = F = 0 \; , \, S = \sigma(\rho)
\end{eqnarray}
there. 
We solve the Einstein equations and gauge condition $\xi^\mu = 0$ as a power series in $z$ near the boundary at $z = 0$, and then transforming to Fefferman-Graham coordinates, identify the boundary metric as that in \eqref{eq:boundarymetric}.
We then extract the 
vev of the dual CFT
stress tensor from the $z^3$ terms in the expansions of $T, V, \ldots, A$ using
holographic renormalisation \cite{deHaro:2000xn}. Details are given in appendix B.

For regularity at the locally AdS boundary we require the reference metric to obey the same boundary conditions as the metric. 
We choose the reference metric 
to be a boosted homogeneous black brane but with $S$ deformed to obey the boundary requirement so,
\begin{eqnarray}
\label{eq:ref}
S &=& \sigma(\rho) \, , \, T = 1 -  c_r^2 \left(z/z_0 \right)^3 \, , \, B = 1 + s_r^2 \left(z/z_0 \right)^3  \\
F &=& - s_r/B \, , \, A = - s_r^2/B  \, , \, V = c_r \, , \, U = s_r c_r \left({z}/{z_0}\right)^3 \nonumber
\end{eqnarray}
for constants $z_0$ 
and $r$, with $c_r = \cosh r$, $s_r = \sinh r$.
The horizon of the reference metric is at $z = z_0$. We emphasise that this is not a solution to the Einstein equations due to the non-trivial $\sigma(\rho)$.

We compactify $\rho$ to a coordinate $x \in [-1, 1]$ where $d \rho = dx / (1-x^2)^2$. In particular this implies that $\rho \sim 1/(x-1)$ as $x \to 1$ and similarly for $x = -1$. Since for a black brane perturbations should decay exponentially in $\rho$ as $\rho \to \pm \infty$ we then expect in our coordinates (i.e. for our reference metric above) all the functions $T, \ldots, S$  
will have all $x$ derivatives vanishing as $x \to \pm 1$.

We work in the coordinate domain $x \in [-1, 1]$ and $z \in [ 0, z_{max} ]$. 
The solutions we find have a future horizon whose position in $z$ is given as a function $z = H(x)$ where $0<H(x)<z_{max}$ so that horizon regularity is imposed
by smoothness of $T, V, \ldots , A$ there.
For $z = z_{max}$ and $-1 < x < -1$ we impose the equations of motion as for the interior points. We emphasise that there is no boundary condition at $z = z_{max}$. Finally at $x = \pm 1$ we impose Neumann boundary conditions for all the functions, as 
$x$ derivatives  should vanish there.

For our solutions we expect two moduli which we can take as the surface gravity and velocity of the horizon in the asymptotic region $\rho \to -\infty$. In the dual picture, these are the temperature and velocity of the inflowing plasma
which is in equilibrium. 
These moduli are not fixed by the reference metric in this ingoing method, and thus we must fix two pieces of information to specify a locally unique solution. This may be done in a myriad of ways, but we have found a numerically robust method is to fix a Dirichlet condition on $V$ at $z = z_{max}$ and $x = -1$ and on $T$ at $z = z_{max}$ and $x = +1$, instead of the Neumann conditions for these at those points. At these two points the value of $V$ at $x = -1$ and $T$ at $x = +1$ are chosen to be those for the reference metric. Thus the two constants $z_0$ and $r$ in the reference metric become the two parameters of the solutions controlling the ingoing plasma 
temperature, $T_0$, and velocity, $v_0$. 

We take the initial guess for the metric to be the reference metric. We use finite differencing to obtain solutions, and discuss the tests of convergence in detail in appendix C, finding our code produces approximately fourth order convergence.
For the resolutions used, up to $70\times280$ in $z$ and $x$, the maximum fractional local error in the Einstein equations   outside the horizon, is better than $\sim 10^{-7}$. Hence these are very good numerical solutions
and we have checked they are indeed Einstein metrics, as we require, rather than Ricci solitons. 
Convergence tests for the extraction of the boundary stress tensor (which depends on multiple derivatives of the metric functions) indicate it is accurate to better than percent level.

\section{From hydrodynamics to quenches}

\begin{figure}
\includegraphics[clip,width=3.5in]{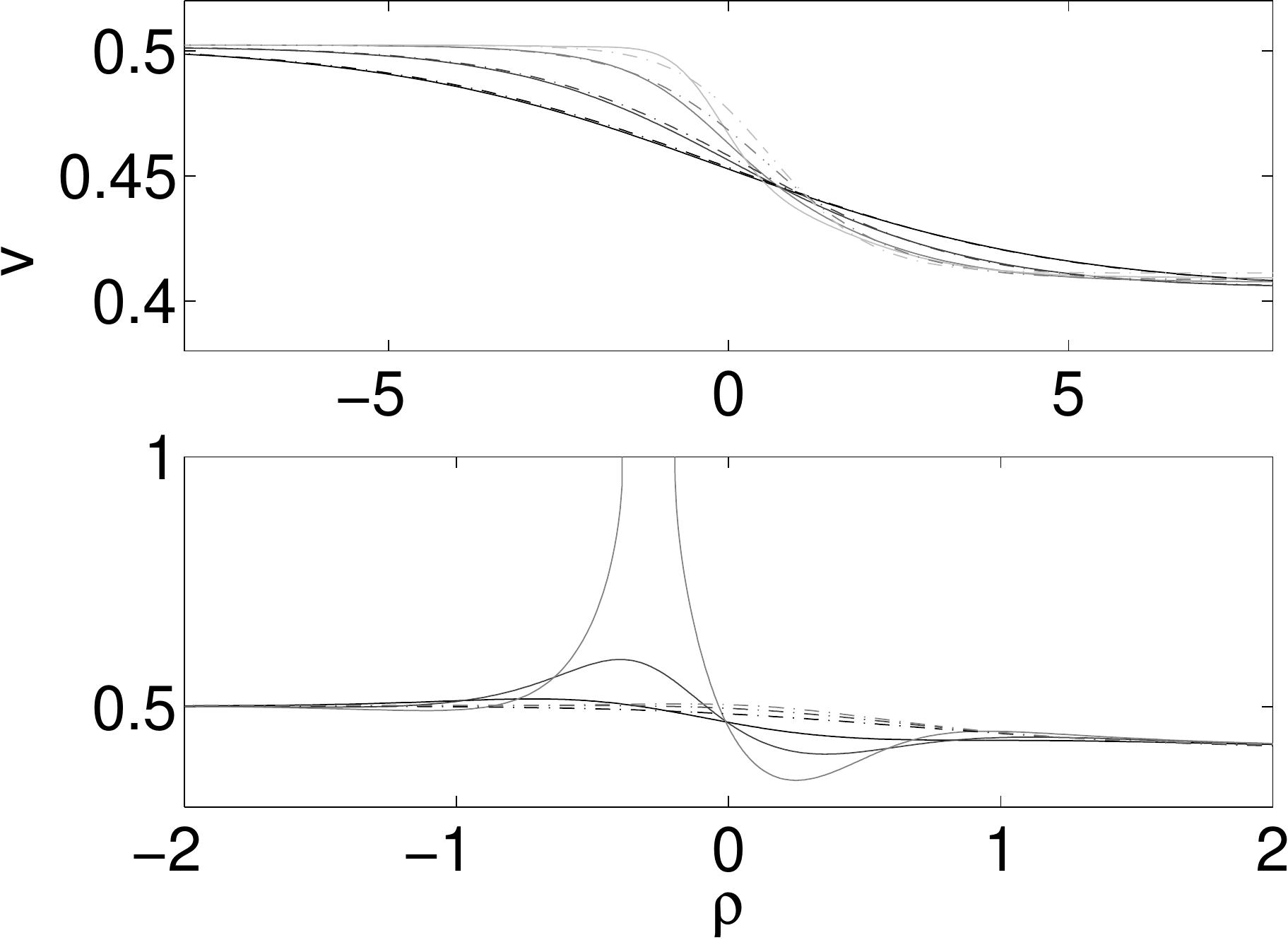} 
  \caption{
  Velocity $v$
   vs. $\rho$ for the flows obtained from the dual gravity solution 
  (solid lines) and by solving the equations of fluid/gravity viscous hydrodynamics (dashed lines),  
  for $\beta = 0.2, 0.3, 0.5, 0.7$ $(\textit{top})$ and $\beta= 1, 1.5, 2$. $(\textit{bottom})$. 
 The plasma flows from left to right, starting in the same homogeneous equilibrium state for the different $\beta$.
  For small $\beta$ we see good agreement with hydrodynamics.
  For large $\beta$ we see strong deviations in the region $| \beta \rho | \lesssim 1$ where the boundary metric varies quickly. As $ \beta \rho  \to + \infty$ the plasma equilibrates and recovers homogeneity. For $\beta = 2.$ we see the flow is  superluminal in the quench region.
  }
  \label{fig:flows}
\end{figure}

We now present data where the ingoing homogeneous plasma has subsonic velocity $v_0 = 0.50$, and temperature $T_0 =  0.24$ in our units. 
Since the boundary theory is a CFT, any other temperature is related by an appropriate scaling, 
and this value is taken for convenience.
We choose the boundary metric to have $\alpha=0.4$, and we adjust $\beta$ to move between a slowly or rapidly varying geometry.
This value of $\alpha$ is sufficiently large that the boundary metric deformation from Minkowski cannot  be described by perturbation theory. As we shall see, the deviation from homogeneous behaviour will correspondingly be large.
With these data we find the dual gravity solution and from it extract the vevs of the CFT stress tensor components $T_{tt}, T_{t\rho}, T_{\rho\rho}$ and $T_{yy}$.
The conservation equation together with tracelessness implies that all the information in the stress tensor is characterised by a single function of $\rho$. We choose to plot the (scale invariant) function defined by,
\begin{eqnarray}
 \frac{v}{1 + v^2} = \frac{ \langle T^{t\rho}\rangle}{ \langle T^{tt} + T^{\rho\rho} \rangle }
\end{eqnarray} 
for $0 \le v \le 1$ where $v$ gives the local velocity of the plasma in the stationary frame.
In figure \ref{fig:flows} we plot this function for various $\beta$ between $0.2$ and $2$.

In this plot we show the same quantity for the fluid/gravity viscous hydrodynamics with the same ingoing flow data - see appendix B for details. 
We see that for the smallest $\beta = 0.2$ the agreement of the gravity stress tensor with that of viscous hydrodynamics is good. The agreement becomes worse as $\beta $ increases and higher derivative terms in the hydrodynamic expansion become important. For $\beta \simeq O(1)$ the hydrodynamic approximation breaks down
and we are in the quench regime. The bulk solutions remains perfectly smooth and allow us to compute the behaviour of this strongly coupled plasma flow. The deviation from hydrodynamics becomes large; for $\beta = 2$ we find the local plasma velocity $v$ becomes superluminal in a region where the metric is curved. This presumably indicates that for sufficient quench strength, $\beta$, these flows become unstable. We note that whilst the stress tensor is superluminal, all its components are well behaved -  for example in appendix B we display $\langle T^{tt}\rangle$.

Interestingly we find the equilibrated outgoing plasma has a temperature, and hence entropy density, that is roughly independent of $\beta$. The same is true for the fluid/gravity viscous hydrodynamics. 
One can see in figure \ref{fig:flows} that the outgoing velocities $v$ are numerically close for the different $\beta$, although they are not obliged to be by stress energy conservation.
We emphasise that whilst the total entropy generated in these flows is similar for different $\beta$, the region where the spacetime is curved and hence this entropy is generated is very different, becoming small for large $\beta$. Hence for strong quenches the entropy density in the plasma is generated in a sudden non-adiabatic manner.

\begin{figure}
  \includegraphics[clip,width=3.5in]{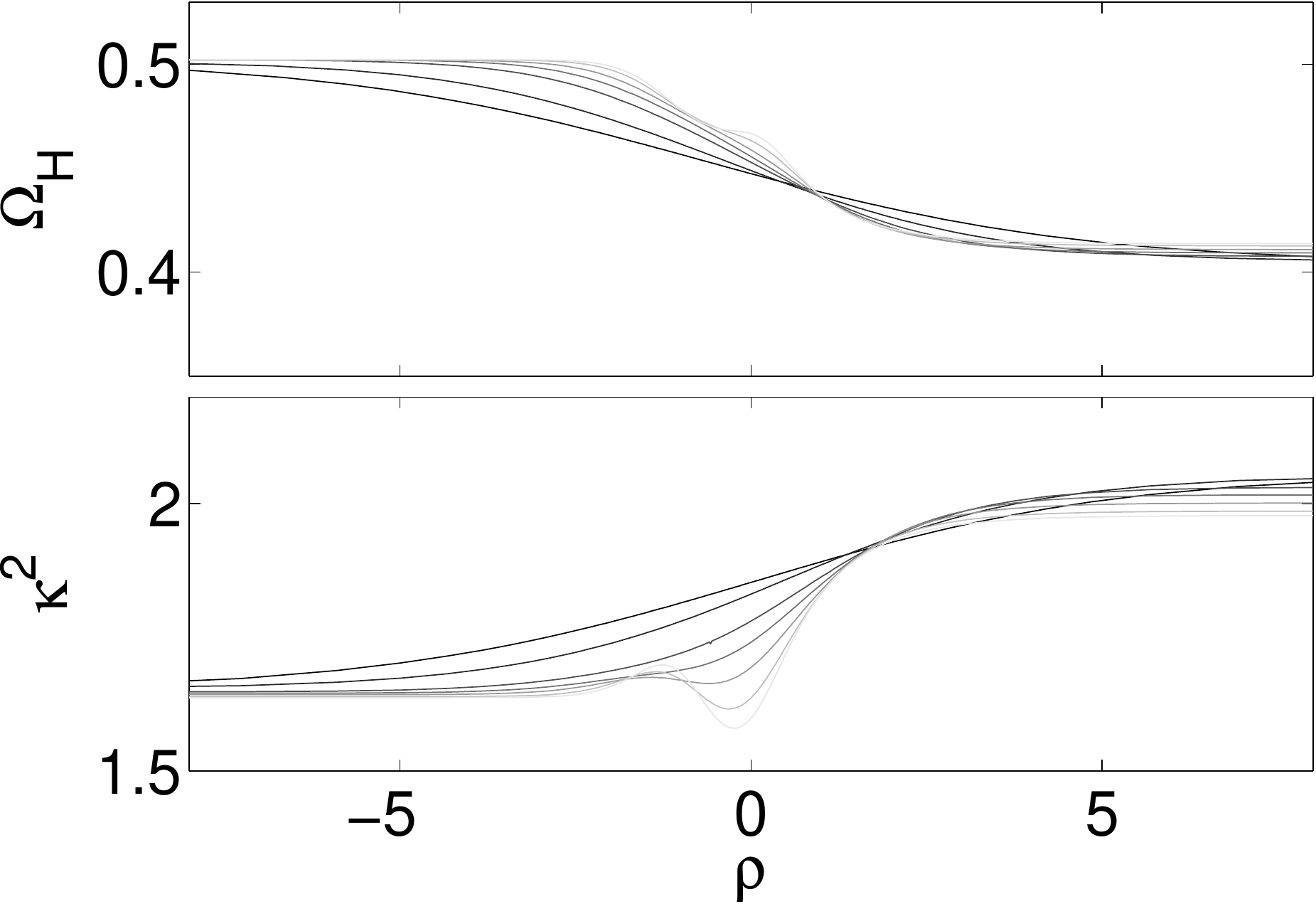}
  \caption{ Velocity of the horizon $\Omega_H$ $(\textit{top})$ and surface gravity $\kappa^2$ $(\textit{bottom})$ as functions of $\rho$ for the same flows as in Fig. \ref{fig:flows}. 
  These functions explicitly depend on $\rho$ as the horizon is non-Killing. 
  Near the asymptotic regions $\rho\to\pm \infty$ both $\Omega_H$ and $\kappa$ become constant since our solutions approach homogeneous boosted black branes.
    }
\label{fig:velocity}
\end{figure}

The horizon, defined by the zero set of $h(z,\rho)=z-H(\rho)$ is a null surface so that,
$
g^{\mu\nu}\,\partial_\mu h\,\partial_\nu h=0
$. This is an o.d.e. for $H(\rho)$ which can be solved to find the horizon location. The null tangent to the horizon can be written,
$
\chi = \frac{\partial}{\partial t}+\Omega_H(\rho) R\,,
$
where $R$ has unit norm $R^2 = 1$, is tangent to the horizon and orthogonal to $\partial/\partial t$ and $\partial/\partial y$. Then $\Omega_H(\rho)$ gives the local velocity of the horizon, and is plotted in figure \ref{fig:velocity}.  
We note this is well behaved even for the flow with $\beta = 2$ which has superluminal boundary stress tensor.
The boundary metric, and consequently the bulk metric, explicitly depend 
on $\rho$ and so $\partial / \partial \rho$ is not Killing. Thus the spacetime motion is not rigid, and hence the local velocity $\Omega_H$ explicitly depends on $\rho$, rather than being constant.
We also compute the surface gravity $\kappa$ defined as $\nabla^\mu ( \chi_\nu \chi^\nu ) =-2 \kappa \chi^\mu$. Again this surface gravity is not constant, and is plotted in the same figure. It is also well behaved for $\beta = 2$.
Further details of the solutions are given in appendices B and C.

\section{Summary}

We have proposed a new numerical method to find stationary black hole solutions that do not have Killing horizons, and hence do not rigidly move, using the harmonic Einstein equations posed on an ingoing domain that pierces the future horizon.
We have explicitly constructed holographic duals to time independent plasma flows in a static geometry that interpolates smoothly between two asymptotic Minkowski regions. For gentle interpolations the plasma behaves as a viscous fluid, as predicted by the fluid/gravity correspondence. 
For sharp interpolations there is no hydrodynamic description, yet the dual black hole allows us to compute 
microscopic properties of this strongly coupled far-from-equilibrium plasma, such as the vev of the stress tensor.
Such solutions are the stationary analog of dynamical quenches. Interestingly we find for a sufficiently strong quench the stress tensor vev may become superluminal in some region. This likely indicates that these flows become unstable for sufficient quench strength. It is possible this instability is turbulent in nature, in analogy with global AdS-Schwarzschild where there is thought to be a turbulent instability for superradiant solutions which correspondingly have superluminal dual plasma  \cite{Dias:2011ss}.
\\

\noindent
{\it Comment: } We understand that \footnote{S.~Fischetti,  D.~Marolf, and J.~Santos, `AdS flowing black funnels:  Stationary AdS black holes with non-Killing horizons and heat transport in the dual CFT' {\it To appear.}} numerically finds flowing funnel solutions which have non-Killing horizons.

\subsection*{Acknowledgments.} We thank Gustav Holzegel, Luis Lehner, Don Marolf, Rob Myers, Jorge Santos and Benson Way for valuable discussions. PF is supported by an EPSRC postdoctoral fellowship [EP/H027106/1].

\bibliography{fluidbib}

\begin{thebibliography}{46}%
\makeatletter
\providecommand \@ifxundefined [1]{%
 \@ifx{#1\undefined}
}%
\providecommand \@ifnum [1]{%
 \ifnum #1\expandafter \@firstoftwo
 \else \expandafter \@secondoftwo
 \fi
}%
\providecommand \@ifx [1]{%
 \ifx #1\expandafter \@firstoftwo
 \else \expandafter \@secondoftwo
 \fi
}%
\providecommand \natexlab [1]{#1}%
\providecommand \enquote  [1]{``#1''}%
\providecommand \bibnamefont  [1]{#1}%
\providecommand \bibfnamefont [1]{#1}%
\providecommand \citenamefont [1]{#1}%
\providecommand \href@noop [0]{\@secondoftwo}%
\providecommand \href [0]{\begingroup \@sanitize@url \@href}%
\providecommand \@href[1]{\@@startlink{#1}\@@href}%
\providecommand \@@href[1]{\endgroup#1\@@endlink}%
\providecommand \@sanitize@url [0]{\catcode `\\12\catcode `\$12\catcode
  `\&12\catcode `\#12\catcode `\^12\catcode `\_12\catcode `\%12\relax}%
\providecommand \@@startlink[1]{}%
\providecommand \@@endlink[0]{}%
\providecommand \url  [0]{\begingroup\@sanitize@url \@url }%
\providecommand \@url [1]{\endgroup\@href {#1}{\urlprefix }}%
\providecommand \urlprefix  [0]{URL }%
\providecommand \Eprint [0]{\href }%
\providecommand \doibase [0]{http://dx.doi.org/}%
\providecommand \selectlanguage [0]{\@gobble}%
\providecommand \bibinfo  [0]{\@secondoftwo}%
\providecommand \bibfield  [0]{\@secondoftwo}%
\providecommand \translation [1]{[#1]}%
\providecommand \BibitemOpen [0]{}%
\providecommand \bibitemStop [0]{}%
\providecommand \bibitemNoStop [0]{.\EOS\space}%
\providecommand \EOS [0]{\spacefactor3000\relax}%
\providecommand \BibitemShut  [1]{\csname bibitem#1\endcsname}%
\let\auto@bib@innerbib\@empty
\bibitem [{\citenamefont {Maldacena}(1998)}]{Maldacena:1997re}%
  \BibitemOpen
  \bibfield  {author} {\bibinfo {author} {\bibfnamefont {J.~M.}\ \bibnamefont
  {Maldacena}},\ }\href@noop {} {\bibfield  {journal} {\bibinfo  {journal}
  {Adv.Theor.Math.Phys.}\ }\textbf {\bibinfo {volume} {2}},\ \bibinfo {pages}
  {231} (\bibinfo {year} {1998})},\ \Eprint
  {http://arxiv.org/abs/hep-th/9711200} {arXiv:hep-th/9711200 [hep-th]}
  \BibitemShut {NoStop}%
\bibitem [{\citenamefont {Gubser}\ \emph {et~al.}(1998)\citenamefont {Gubser},
  \citenamefont {Klebanov},\ and\ \citenamefont {Polyakov}}]{Gubser:GTC}%
  \BibitemOpen
  \bibfield  {author} {\bibinfo {author} {\bibfnamefont {S.~S.}\ \bibnamefont
  {Gubser}}, \bibinfo {author} {\bibfnamefont {I.~R.}\ \bibnamefont
  {Klebanov}}, \ and\ \bibinfo {author} {\bibfnamefont {A.~M.}\ \bibnamefont
  {Polyakov}},\ }\href {\doibase DOI:10.1016/S0370-2693(98)00377-3} {\bibfield
  {journal} {\bibinfo  {journal} {Phys. Lett. B}\ }\textbf {\bibinfo {volume}
  {428}},\ \bibinfo {pages} {105} (\bibinfo {year} {1998})}\BibitemShut
  {NoStop}%
\bibitem [{\citenamefont {Witten}(1998)}]{Witten:AdSholo}%
  \BibitemOpen
  \bibfield  {author} {\bibinfo {author} {\bibfnamefont {E.}~\bibnamefont
  {Witten}},\ }\href@noop {} {\bibfield  {journal} {\bibinfo  {journal} {Adv.
  Theor. Math. Phys}\ }\textbf {\bibinfo {volume} {2}},\ \bibinfo {pages} {253}
  (\bibinfo {year} {1998})}\BibitemShut {NoStop}%
\bibitem [{\citenamefont {Kovtun}\ \emph {et~al.}(2003)\citenamefont {Kovtun},
  \citenamefont {Son},\ and\ \citenamefont {Starinets}}]{Kovtun:2003wp}%
  \BibitemOpen
  \bibfield  {author} {\bibinfo {author} {\bibfnamefont {P.}~\bibnamefont
  {Kovtun}}, \bibinfo {author} {\bibfnamefont {D.~T.}\ \bibnamefont {Son}}, \
  and\ \bibinfo {author} {\bibfnamefont {A.~O.}\ \bibnamefont {Starinets}},\
  }\href@noop {} {\bibfield  {journal} {\bibinfo  {journal} {JHEP}\ }\textbf
  {\bibinfo {volume} {0310}},\ \bibinfo {pages} {064} (\bibinfo {year}
  {2003})},\ \Eprint {http://arxiv.org/abs/hep-th/0309213}
  {arXiv:hep-th/0309213 [hep-th]} \BibitemShut {NoStop}%
\bibitem [{\citenamefont {Buchel}\ and\ \citenamefont
  {Liu}(2004)}]{Buchel:2003tz}%
  \BibitemOpen
  \bibfield  {author} {\bibinfo {author} {\bibfnamefont {A.}~\bibnamefont
  {Buchel}}\ and\ \bibinfo {author} {\bibfnamefont {J.~T.}\ \bibnamefont
  {Liu}},\ }\href {\doibase 10.1103/PhysRevLett.93.090602} {\bibfield
  {journal} {\bibinfo  {journal} {Phys.Rev.Lett.}\ }\textbf {\bibinfo {volume}
  {93}},\ \bibinfo {pages} {090602} (\bibinfo {year} {2004})},\ \Eprint
  {http://arxiv.org/abs/hep-th/0311175} {arXiv:hep-th/0311175 [hep-th]}
  \BibitemShut {NoStop}%
\bibitem [{\citenamefont {Kovtun}\ \emph {et~al.}(2005)\citenamefont {Kovtun},
  \citenamefont {Son},\ and\ \citenamefont {Starinets}}]{Kovtun:2004de}%
  \BibitemOpen
  \bibfield  {author} {\bibinfo {author} {\bibfnamefont {P.}~\bibnamefont
  {Kovtun}}, \bibinfo {author} {\bibfnamefont {D.}~\bibnamefont {Son}}, \ and\
  \bibinfo {author} {\bibfnamefont {A.}~\bibnamefont {Starinets}},\ }\href
  {\doibase 10.1103/PhysRevLett.94.111601} {\bibfield  {journal} {\bibinfo
  {journal} {Phys.Rev.Lett.}\ }\textbf {\bibinfo {volume} {94}},\ \bibinfo
  {pages} {111601} (\bibinfo {year} {2005})},\ \Eprint
  {http://arxiv.org/abs/hep-th/0405231} {arXiv:hep-th/0405231 [hep-th]}
  \BibitemShut {NoStop}%
\bibitem [{\citenamefont {Baier}\ \emph {et~al.}(2008)\citenamefont {Baier},
  \citenamefont {Romatschke}, \citenamefont {Son}, \citenamefont {Starinets},\
  and\ \citenamefont {Stephanov}}]{Baier:2007ix}%
  \BibitemOpen
  \bibfield  {author} {\bibinfo {author} {\bibfnamefont {R.}~\bibnamefont
  {Baier}}, \bibinfo {author} {\bibfnamefont {P.}~\bibnamefont {Romatschke}},
  \bibinfo {author} {\bibfnamefont {D.~T.}\ \bibnamefont {Son}}, \bibinfo
  {author} {\bibfnamefont {A.~O.}\ \bibnamefont {Starinets}}, \ and\ \bibinfo
  {author} {\bibfnamefont {M.~A.}\ \bibnamefont {Stephanov}},\ }\href {\doibase
  10.1088/1126-6708/2008/04/100} {\bibfield  {journal} {\bibinfo  {journal}
  {JHEP}\ }\textbf {\bibinfo {volume} {0804}},\ \bibinfo {pages} {100}
  (\bibinfo {year} {2008})},\ \Eprint {http://arxiv.org/abs/0712.2451}
  {arXiv:0712.2451 [hep-th]} \BibitemShut {NoStop}%
\bibitem [{\citenamefont {Bhattacharyya}\ \emph
  {et~al.}(2008{\natexlab{a}})\citenamefont {Bhattacharyya}, \citenamefont
  {Hubeny}, \citenamefont {Minwalla},\ and\ \citenamefont
  {Rangamani}}]{Bhattacharyya:2008jc}%
  \BibitemOpen
  \bibfield  {author} {\bibinfo {author} {\bibfnamefont {S.}~\bibnamefont
  {Bhattacharyya}}, \bibinfo {author} {\bibfnamefont {V.~E.}\ \bibnamefont
  {Hubeny}}, \bibinfo {author} {\bibfnamefont {S.}~\bibnamefont {Minwalla}}, \
  and\ \bibinfo {author} {\bibfnamefont {M.}~\bibnamefont {Rangamani}},\ }\href
  {\doibase 10.1088/1126-6708/2008/02/045} {\bibfield  {journal} {\bibinfo
  {journal} {JHEP}\ }\textbf {\bibinfo {volume} {0802}},\ \bibinfo {pages}
  {045} (\bibinfo {year} {2008}{\natexlab{a}})},\ \Eprint
  {http://arxiv.org/abs/0712.2456} {arXiv:0712.2456 [hep-th]} \BibitemShut
  {NoStop}%
\bibitem [{\citenamefont {Hartnoll}(2009)}]{Hartnoll:Lectures}%
  \BibitemOpen
  \bibfield  {author} {\bibinfo {author} {\bibfnamefont {S.~A.}\ \bibnamefont
  {Hartnoll}},\ }\href@noop {} {\bibfield  {journal} {\bibinfo  {journal}
  {Class. Quant. Grav.}\ }\textbf {\bibinfo {volume} {26}},\ \bibinfo {pages}
  {224002} (\bibinfo {year} {2009})}\BibitemShut {NoStop}%
\bibitem [{\citenamefont {McGreevy}(2010)}]{McGreevy:Notes}%
  \BibitemOpen
  \bibfield  {author} {\bibinfo {author} {\bibfnamefont {J.}~\bibnamefont
  {McGreevy}},\ }\href {\doibase doi:10.1155/2010/723105} {\bibfield  {journal}
  {\bibinfo  {journal} {Adv. High Energy Phys.}\ ,\ \bibinfo {pages} {723105}}
  (\bibinfo {year} {2010})}\BibitemShut {NoStop}%
\bibitem [{\citenamefont {Danielsson}\ \emph {et~al.}(1999)\citenamefont
  {Danielsson}, \citenamefont {Keski-Vakkuri},\ and\ \citenamefont
  {Kruczenski}}]{Danielsson:1999zt}%
  \BibitemOpen
  \bibfield  {author} {\bibinfo {author} {\bibfnamefont {U.~H.}\ \bibnamefont
  {Danielsson}}, \bibinfo {author} {\bibfnamefont {E.}~\bibnamefont
  {Keski-Vakkuri}}, \ and\ \bibinfo {author} {\bibfnamefont {M.}~\bibnamefont
  {Kruczenski}},\ }\href {\doibase 10.1016/S0550-3213(99)00511-8} {\bibfield
  {journal} {\bibinfo  {journal} {Nucl. Phys. B}\ }\textbf {\bibinfo {volume}
  {563}},\ \bibinfo {pages} {279} (\bibinfo {year} {1999})}\BibitemShut
  {NoStop}%
\bibitem [{\citenamefont {Giddings}\ and\ \citenamefont
  {Ross}(2000)}]{Giddings:1999zu}%
  \BibitemOpen
  \bibfield  {author} {\bibinfo {author} {\bibfnamefont {S.~B.}\ \bibnamefont
  {Giddings}}\ and\ \bibinfo {author} {\bibfnamefont {S.~F.}\ \bibnamefont
  {Ross}},\ }\href {\doibase 10.1103/PhysRevD.61.024036} {\bibfield  {journal}
  {\bibinfo  {journal} {Phys. Rev. D}\ }\textbf {\bibinfo {volume} {61}},\
  \bibinfo {pages} {024036} (\bibinfo {year} {2000})}\BibitemShut {NoStop}%
\bibitem [{\citenamefont {Bhattacharyya}\ and\ \citenamefont
  {Minwalla}(2009)}]{Bhattacharyya:2009uu}%
  \BibitemOpen
  \bibfield  {author} {\bibinfo {author} {\bibfnamefont {S.}~\bibnamefont
  {Bhattacharyya}}\ and\ \bibinfo {author} {\bibfnamefont {S.}~\bibnamefont
  {Minwalla}},\ }\href {\doibase 10.1088/1126-6708/2009/09/034} {\bibfield
  {journal} {\bibinfo  {journal} {JHEP}\ }\textbf {\bibinfo {volume} {0909}},\
  \bibinfo {pages} {034} (\bibinfo {year} {2009})}\BibitemShut {NoStop}%
\bibitem [{\citenamefont {Das}\ \emph {et~al.}(2010)\citenamefont {Das},
  \citenamefont {Nishioka},\ and\ \citenamefont {Takayanagi}}]{Das:2010yw}%
  \BibitemOpen
  \bibfield  {author} {\bibinfo {author} {\bibfnamefont {S.~R.}\ \bibnamefont
  {Das}}, \bibinfo {author} {\bibfnamefont {T.}~\bibnamefont {Nishioka}}, \
  and\ \bibinfo {author} {\bibfnamefont {T.}~\bibnamefont {Takayanagi}},\
  }\href {\doibase 10.1007/JHEP07(2010)071} {\bibfield  {journal} {\bibinfo
  {journal} {JHEP}\ }\textbf {\bibinfo {volume} {1007}},\ \bibinfo {pages}
  {071} (\bibinfo {year} {2010})}\BibitemShut {NoStop}%
\bibitem [{\citenamefont {Albash}\ and\ \citenamefont
  {Johnson}(2011)}]{Albash:2010mv}%
  \BibitemOpen
  \bibfield  {author} {\bibinfo {author} {\bibfnamefont {T.}~\bibnamefont
  {Albash}}\ and\ \bibinfo {author} {\bibfnamefont {C.~V.}\ \bibnamefont
  {Johnson}},\ }\href {\doibase 10.1088/1367-2630/13/4/045017} {\bibfield
  {journal} {\bibinfo  {journal} {New J. Phys.}\ }\textbf {\bibinfo {volume}
  {13}},\ \bibinfo {pages} {045017} (\bibinfo {year} {2011})}\BibitemShut
  {NoStop}%
\bibitem [{\citenamefont {Chesler}\ and\ \citenamefont
  {Yaffe}(2009)}]{Chesler:2008hg}%
  \BibitemOpen
  \bibfield  {author} {\bibinfo {author} {\bibfnamefont {P.~M.}\ \bibnamefont
  {Chesler}}\ and\ \bibinfo {author} {\bibfnamefont {L.~G.}\ \bibnamefont
  {Yaffe}},\ }\href {\doibase 10.1103/PhysRevLett.102.211601} {\bibfield
  {journal} {\bibinfo  {journal} {Phys. Rev. Lett.}\ }\textbf {\bibinfo
  {volume} {102}},\ \bibinfo {pages} {211601} (\bibinfo {year}
  {2009})}\BibitemShut {NoStop}%
\bibitem [{\citenamefont {Murata}\ \emph {et~al.}(2010)\citenamefont {Murata},
  \citenamefont {Kinoshita},\ and\ \citenamefont {Tanahashi}}]{Murata:Noneq}%
  \BibitemOpen
  \bibfield  {author} {\bibinfo {author} {\bibfnamefont {K.}~\bibnamefont
  {Murata}}, \bibinfo {author} {\bibfnamefont {S.}~\bibnamefont {Kinoshita}}, \
  and\ \bibinfo {author} {\bibfnamefont {N.}~\bibnamefont {Tanahashi}},\ }\href
  {\doibase 10.1007/JHEP07(2010)050} {\bibfield  {journal} {\bibinfo  {journal}
  {JHEP}\ }\textbf {\bibinfo {volume} {1007}},\ \bibinfo {pages} {050}
  (\bibinfo {year} {2010})}\BibitemShut {NoStop}%
\bibitem [{\citenamefont {Bizon}\ and\ \citenamefont
  {Rostworowski}(2011)}]{Bizon:2011gg}%
  \BibitemOpen
  \bibfield  {author} {\bibinfo {author} {\bibfnamefont {P.}~\bibnamefont
  {Bizon}}\ and\ \bibinfo {author} {\bibfnamefont {A.}~\bibnamefont
  {Rostworowski}},\ }\href {\doibase 10.1103/PhysRevLett.107.031102} {\bibfield
   {journal} {\bibinfo  {journal} {Phys. Rev. Lett.}\ }\textbf {\bibinfo
  {volume} {107}},\ \bibinfo {pages} {031102} (\bibinfo {year}
  {2011})}\BibitemShut {NoStop}%
\bibitem [{\citenamefont {Garfinkle}\ and\ \citenamefont
  {Pando~Zayas}(2011)}]{Garfinkle:2011hm}%
  \BibitemOpen
  \bibfield  {author} {\bibinfo {author} {\bibfnamefont {D.}~\bibnamefont
  {Garfinkle}}\ and\ \bibinfo {author} {\bibfnamefont {L.~A.}\ \bibnamefont
  {Pando~Zayas}},\ }\href {\doibase 10.1103/PhysRevD.84.066006} {\bibfield
  {journal} {\bibinfo  {journal} {Phys. Rev. D}\ }\textbf {\bibinfo {volume}
  {84}},\ \bibinfo {pages} {066006} (\bibinfo {year} {2011})}\BibitemShut
  {NoStop}%
\bibitem [{\citenamefont {Bantilan}\ \emph {et~al.}(2012)\citenamefont
  {Bantilan}, \citenamefont {Pretorius},\ and\ \citenamefont
  {Gubser}}]{Bantilan:2012vu}%
  \BibitemOpen
  \bibfield  {author} {\bibinfo {author} {\bibfnamefont {H.}~\bibnamefont
  {Bantilan}}, \bibinfo {author} {\bibfnamefont {F.}~\bibnamefont {Pretorius}},
  \ and\ \bibinfo {author} {\bibfnamefont {S.~S.}\ \bibnamefont {Gubser}},\
  }\href {\doibase 10.1103/PhysRevD.85.084038} {\bibfield  {journal} {\bibinfo
  {journal} {Phys. Rev. D}\ }\textbf {\bibinfo {volume} {85}},\ \bibinfo
  {pages} {084038} (\bibinfo {year} {2012})}\BibitemShut {NoStop}%
\bibitem [{\citenamefont {Buchel}\ \emph {et~al.}()\citenamefont {Buchel},
  \citenamefont {Lehner},\ and\ \citenamefont {Myers}}]{Buchel:Thermal}%
  \BibitemOpen
  \bibfield  {author} {\bibinfo {author} {\bibfnamefont {A.}~\bibnamefont
  {Buchel}}, \bibinfo {author} {\bibfnamefont {L.}~\bibnamefont {Lehner}}, \
  and\ \bibinfo {author} {\bibfnamefont {R.~C.}\ \bibnamefont {Myers}},\
  }\href@noop {} {\ }\Eprint {http://arxiv.org/abs/1206.6785} {arXiv:1206.6785}
  \BibitemShut {NoStop}%
\bibitem [{\citenamefont {Bhaseen}\ \emph {et~al.}(2012)\citenamefont
  {Bhaseen}, \citenamefont {Gauntlett}, \citenamefont {Simons}, \citenamefont
  {Sonner},\ and\ \citenamefont {Wiseman}}]{Bhaseen:2012gg}%
  \BibitemOpen
  \bibfield  {author} {\bibinfo {author} {\bibfnamefont {M.}~\bibnamefont
  {Bhaseen}}, \bibinfo {author} {\bibfnamefont {J.~P.}\ \bibnamefont
  {Gauntlett}}, \bibinfo {author} {\bibfnamefont {B.}~\bibnamefont {Simons}},
  \bibinfo {author} {\bibfnamefont {J.}~\bibnamefont {Sonner}}, \ and\ \bibinfo
  {author} {\bibfnamefont {T.}~\bibnamefont {Wiseman}},\ }\href@noop {} {\
  (\bibinfo {year} {2012})},\ \Eprint {http://arxiv.org/abs/1207.4194}
  {arXiv:1207.4194 [hep-th]} \BibitemShut {NoStop}%
\bibitem [{\citenamefont {Bhattacharyya}\ \emph {et~al.}(2009)\citenamefont
  {Bhattacharyya}, \citenamefont {Loganayagam}, \citenamefont {Minwalla},
  \citenamefont {Nampuri}, \citenamefont {Trivedi} \emph
  {et~al.}}]{Bhattacharyya:2008ji}%
  \BibitemOpen
  \bibfield  {author} {\bibinfo {author} {\bibfnamefont {S.}~\bibnamefont
  {Bhattacharyya}}, \bibinfo {author} {\bibfnamefont {R.}~\bibnamefont
  {Loganayagam}}, \bibinfo {author} {\bibfnamefont {S.}~\bibnamefont
  {Minwalla}}, \bibinfo {author} {\bibfnamefont {S.}~\bibnamefont {Nampuri}},
  \bibinfo {author} {\bibfnamefont {S.~P.}\ \bibnamefont {Trivedi}},  \emph
  {et~al.},\ }\href {\doibase 10.1088/1126-6708/2009/02/018} {\bibfield
  {journal} {\bibinfo  {journal} {JHEP}\ }\textbf {\bibinfo {volume} {0902}},\
  \bibinfo {pages} {018} (\bibinfo {year} {2009})},\ \Eprint
  {http://arxiv.org/abs/0806.0006} {arXiv:0806.0006 [hep-th]} \BibitemShut
  {NoStop}%
\bibitem [{Note1()}]{Note1}%
  \BibitemOpen
  \bibinfo {note} {It is interesting to contrast this with the solutions of
  \cite {Dias:2011at} which have a Killing horizon, but are neither stationary
  nor axisymmetric if one considers the metric \protect \emph {and} matter.
  However, considering the metric alone, the solution is stationary and does
  rigidly rotate. We refer to rigidity with reference to the metric alone in
  this work.}\BibitemShut {Stop}%
\bibitem [{\citenamefont {Hawking}\ and\ \citenamefont
  {Ellis}(1973)}]{Hawking:1973uf}%
  \BibitemOpen
  \bibfield  {author} {\bibinfo {author} {\bibfnamefont {S.}~\bibnamefont
  {Hawking}}\ and\ \bibinfo {author} {\bibfnamefont {G.}~\bibnamefont
  {Ellis}},\ }\href@noop {} {\  (\bibinfo {year} {1973})}\BibitemShut {NoStop}%
\bibitem [{\citenamefont {Hollands}\ \emph {et~al.}(2007)\citenamefont
  {Hollands}, \citenamefont {Ishibashi},\ and\ \citenamefont
  {Wald}}]{Hollands:2006rj}%
  \BibitemOpen
  \bibfield  {author} {\bibinfo {author} {\bibfnamefont {S.}~\bibnamefont
  {Hollands}}, \bibinfo {author} {\bibfnamefont {A.}~\bibnamefont {Ishibashi}},
  \ and\ \bibinfo {author} {\bibfnamefont {R.~M.}\ \bibnamefont {Wald}},\
  }\href {\doibase 10.1007/s00220-007-0216-4} {\bibfield  {journal} {\bibinfo
  {journal} {Commun.Math.Phys.}\ }\textbf {\bibinfo {volume} {271}},\ \bibinfo
  {pages} {699} (\bibinfo {year} {2007})},\ \Eprint
  {http://arxiv.org/abs/gr-qc/0605106} {arXiv:gr-qc/0605106 [gr-qc]}
  \BibitemShut {NoStop}%
\bibitem [{\citenamefont {Moncrief}\ and\ \citenamefont
  {Isenberg}(2008)}]{Moncrief:2008mr}%
  \BibitemOpen
  \bibfield  {author} {\bibinfo {author} {\bibfnamefont {V.}~\bibnamefont
  {Moncrief}}\ and\ \bibinfo {author} {\bibfnamefont {J.}~\bibnamefont
  {Isenberg}},\ }\href {\doibase 10.1088/0264-9381/25/19/195015} {\bibfield
  {journal} {\bibinfo  {journal} {Class.Quant.Grav.}\ }\textbf {\bibinfo
  {volume} {25}},\ \bibinfo {pages} {195015} (\bibinfo {year} {2008})},\
  \Eprint {http://arxiv.org/abs/0805.1451} {arXiv:0805.1451 [gr-qc]}
  \BibitemShut {NoStop}%
\bibitem [{\citenamefont {Hubeny}\ \emph {et~al.}(2010)\citenamefont {Hubeny},
  \citenamefont {Marolf},\ and\ \citenamefont {Rangamani}}]{Hubeny:2009ru}%
  \BibitemOpen
  \bibfield  {author} {\bibinfo {author} {\bibfnamefont {V.~E.}\ \bibnamefont
  {Hubeny}}, \bibinfo {author} {\bibfnamefont {D.}~\bibnamefont {Marolf}}, \
  and\ \bibinfo {author} {\bibfnamefont {M.}~\bibnamefont {Rangamani}},\ }\href
  {\doibase 10.1088/0264-9381/27/9/095015} {\bibfield  {journal} {\bibinfo
  {journal} {Class.Quant.Grav.}\ }\textbf {\bibinfo {volume} {27}},\ \bibinfo
  {pages} {095015} (\bibinfo {year} {2010})},\ \Eprint
  {http://arxiv.org/abs/0908.2270} {arXiv:0908.2270 [hep-th]} \BibitemShut
  {NoStop}%
\bibitem [{\citenamefont {Fischetti}\ and\ \citenamefont
  {Marolf}(2012)}]{Fischetti:2012ps}%
  \BibitemOpen
  \bibfield  {author} {\bibinfo {author} {\bibfnamefont {S.}~\bibnamefont
  {Fischetti}}\ and\ \bibinfo {author} {\bibfnamefont {D.}~\bibnamefont
  {Marolf}},\ }\href {\doibase 10.1088/0264-9381/29/10/105004} {\bibfield
  {journal} {\bibinfo  {journal} {Class.Quant.Grav.}\ }\textbf {\bibinfo
  {volume} {29}},\ \bibinfo {pages} {105004} (\bibinfo {year} {2012})},\
  \Eprint {http://arxiv.org/abs/1202.5069} {arXiv:1202.5069 [hep-th]}
  \BibitemShut {NoStop}%
\bibitem [{\citenamefont {Khlebnikov}\ \emph {et~al.}(2010)\citenamefont
  {Khlebnikov}, \citenamefont {Kruczenski},\ and\ \citenamefont
  {Michalogiorgakis}}]{Khlebnikov:2010yt}%
  \BibitemOpen
  \bibfield  {author} {\bibinfo {author} {\bibfnamefont {S.}~\bibnamefont
  {Khlebnikov}}, \bibinfo {author} {\bibfnamefont {M.}~\bibnamefont
  {Kruczenski}}, \ and\ \bibinfo {author} {\bibfnamefont {G.}~\bibnamefont
  {Michalogiorgakis}},\ }\href {\doibase 10.1103/PhysRevD.82.125003} {\bibfield
   {journal} {\bibinfo  {journal} {Phys.Rev.}\ }\textbf {\bibinfo {volume}
  {D82}},\ \bibinfo {pages} {125003} (\bibinfo {year} {2010})},\ \Eprint
  {http://arxiv.org/abs/1004.3803} {arXiv:1004.3803 [hep-th]} \BibitemShut
  {NoStop}%
\bibitem [{\citenamefont {Khlebnikov}\ \emph {et~al.}(2011)\citenamefont
  {Khlebnikov}, \citenamefont {Kruczenski},\ and\ \citenamefont
  {Michalogiorgakis}}]{Khlebnikov:2011ka}%
  \BibitemOpen
  \bibfield  {author} {\bibinfo {author} {\bibfnamefont {S.}~\bibnamefont
  {Khlebnikov}}, \bibinfo {author} {\bibfnamefont {M.}~\bibnamefont
  {Kruczenski}}, \ and\ \bibinfo {author} {\bibfnamefont {G.}~\bibnamefont
  {Michalogiorgakis}},\ }\href {\doibase 10.1007/JHEP07(2011)097} {\bibfield
  {journal} {\bibinfo  {journal} {JHEP}\ }\textbf {\bibinfo {volume} {1107}},\
  \bibinfo {pages} {097} (\bibinfo {year} {2011})},\ \Eprint
  {http://arxiv.org/abs/1105.1355} {arXiv:1105.1355 [hep-th]} \BibitemShut
  {NoStop}%
\bibitem [{\citenamefont {Figueras}\ \emph {et~al.}(2011)\citenamefont
  {Figueras}, \citenamefont {Lucietti},\ and\ \citenamefont
  {Wiseman}}]{Figueras:2011va}%
  \BibitemOpen
  \bibfield  {author} {\bibinfo {author} {\bibfnamefont {P.}~\bibnamefont
  {Figueras}}, \bibinfo {author} {\bibfnamefont {J.}~\bibnamefont {Lucietti}},
  \ and\ \bibinfo {author} {\bibfnamefont {T.}~\bibnamefont {Wiseman}},\ }\href
  {\doibase 10.1088/0264-9381/28/21/215018} {\bibfield  {journal} {\bibinfo
  {journal} {Class.Quant.Grav.}\ }\textbf {\bibinfo {volume} {28}},\ \bibinfo
  {pages} {215018} (\bibinfo {year} {2011})},\ \Eprint
  {http://arxiv.org/abs/1104.4489} {arXiv:1104.4489 [hep-th]} \BibitemShut
  {NoStop}%
\bibitem [{\citenamefont {Santos}\ and\ \citenamefont
  {Way}(2012)}]{Santos:2012he}%
  \BibitemOpen
  \bibfield  {author} {\bibinfo {author} {\bibfnamefont {J.~E.}\ \bibnamefont
  {Santos}}\ and\ \bibinfo {author} {\bibfnamefont {B.}~\bibnamefont {Way}},\
  }\href@noop {} {\  (\bibinfo {year} {2012})},\ \Eprint
  {http://arxiv.org/abs/1208.6291} {arXiv:1208.6291 [hep-th]} \BibitemShut
  {NoStop}%
\bibitem [{\citenamefont {Headrick}\ \emph {et~al.}(2010)\citenamefont
  {Headrick}, \citenamefont {Kitchen},\ and\ \citenamefont
  {Wiseman}}]{Headrick:2009pv}%
  \BibitemOpen
  \bibfield  {author} {\bibinfo {author} {\bibfnamefont {M.}~\bibnamefont
  {Headrick}}, \bibinfo {author} {\bibfnamefont {S.}~\bibnamefont {Kitchen}}, \
  and\ \bibinfo {author} {\bibfnamefont {T.}~\bibnamefont {Wiseman}},\ }\href
  {\doibase 10.1088/0264-9381/27/3/035002} {\bibfield  {journal} {\bibinfo
  {journal} {Class.Quant.Grav.}\ }\textbf {\bibinfo {volume} {27}},\ \bibinfo
  {pages} {035002} (\bibinfo {year} {2010})},\ \Eprint
  {http://arxiv.org/abs/0905.1822} {arXiv:0905.1822 [gr-qc]} \BibitemShut
  {NoStop}%
\bibitem [{\citenamefont {Adam}\ \emph {et~al.}(2012)\citenamefont {Adam},
  \citenamefont {Kitchen},\ and\ \citenamefont {Wiseman}}]{Adam:2011dn}%
  \BibitemOpen
  \bibfield  {author} {\bibinfo {author} {\bibfnamefont {A.}~\bibnamefont
  {Adam}}, \bibinfo {author} {\bibfnamefont {S.}~\bibnamefont {Kitchen}}, \
  and\ \bibinfo {author} {\bibfnamefont {T.}~\bibnamefont {Wiseman}},\ }\href
  {\doibase 10.1088/0264-9381/29/16/165002} {\bibfield  {journal} {\bibinfo
  {journal} {Class.Quant.Grav.}\ }\textbf {\bibinfo {volume} {29}},\ \bibinfo
  {pages} {165002} (\bibinfo {year} {2012})},\ \Eprint
  {http://arxiv.org/abs/1105.6347} {arXiv:1105.6347 [gr-qc]} \BibitemShut
  {NoStop}%
\bibitem [{\citenamefont {Wiseman}(2011)}]{Wiseman:2011by}%
  \BibitemOpen
  \bibfield  {author} {\bibinfo {author} {\bibfnamefont {T.}~\bibnamefont
  {Wiseman}},\ }\href@noop {} {\  (\bibinfo {year} {2011})},\ \Eprint
  {http://arxiv.org/abs/1107.5513} {arXiv:1107.5513 [gr-qc]} \BibitemShut
  {NoStop}%
\bibitem [{\citenamefont {Garfinkle}(2002)}]{Garfinkle:2001ni}%
  \BibitemOpen
  \bibfield  {author} {\bibinfo {author} {\bibfnamefont {D.}~\bibnamefont
  {Garfinkle}},\ }\href {\doibase 10.1103/PhysRevD.65.044029} {\bibfield
  {journal} {\bibinfo  {journal} {Phys.Rev.}\ }\textbf {\bibinfo {volume}
  {D65}},\ \bibinfo {pages} {044029} (\bibinfo {year} {2002})},\ \Eprint
  {http://arxiv.org/abs/gr-qc/0110013} {arXiv:gr-qc/0110013 [gr-qc]}
  \BibitemShut {NoStop}%
\bibitem [{\citenamefont {Hollands}\ and\ \citenamefont
  {Yazadjiev}(2008)}]{Hollands:2007aj}%
  \BibitemOpen
  \bibfield  {author} {\bibinfo {author} {\bibfnamefont {S.}~\bibnamefont
  {Hollands}}\ and\ \bibinfo {author} {\bibfnamefont {S.}~\bibnamefont
  {Yazadjiev}},\ }\href {\doibase 10.1007/s00220-008-0516-3} {\bibfield
  {journal} {\bibinfo  {journal} {Commun.Math.Phys.}\ }\textbf {\bibinfo
  {volume} {283}},\ \bibinfo {pages} {749} (\bibinfo {year} {2008})},\ \Eprint
  {http://arxiv.org/abs/0707.2775} {arXiv:0707.2775 [gr-qc]} \BibitemShut
  {NoStop}%
\bibitem [{\citenamefont {Harmark}(2009)}]{Harmark:2009dh}%
  \BibitemOpen
  \bibfield  {author} {\bibinfo {author} {\bibfnamefont {T.}~\bibnamefont
  {Harmark}},\ }\href {\doibase 10.1103/PhysRevD.80.024019} {\bibfield
  {journal} {\bibinfo  {journal} {Phys.Rev.}\ }\textbf {\bibinfo {volume}
  {D80}},\ \bibinfo {pages} {024019} (\bibinfo {year} {2009})},\ \Eprint
  {http://arxiv.org/abs/0904.4246} {arXiv:0904.4246 [hep-th]} \BibitemShut
  {NoStop}%
\bibitem [{Note2()}]{Note2}%
  \BibitemOpen
  \bibinfo {note} {This is to be contrasted with the canonical mixed hyperbolic
  elliptic equation, the Tricomi equation, $\left (\partial ^2 / \partial y^2 +
  y \protect \tmspace +\thinmuskip {.1667em} \partial ^2 / \partial x^2 \right
  ) f = 0$ where smoothness at $y = 0$ (the analog of the horizon or
  ergosurface) imposes no condition on the solution.}\BibitemShut {Stop}%
\bibitem [{\citenamefont {de~Haro}\ \emph {et~al.}(2001)\citenamefont
  {de~Haro}, \citenamefont {Solodukhin},\ and\ \citenamefont
  {Skenderis}}]{deHaro:2000xn}%
  \BibitemOpen
  \bibfield  {author} {\bibinfo {author} {\bibfnamefont {S.}~\bibnamefont
  {de~Haro}}, \bibinfo {author} {\bibfnamefont {S.~N.}\ \bibnamefont
  {Solodukhin}}, \ and\ \bibinfo {author} {\bibfnamefont {K.}~\bibnamefont
  {Skenderis}},\ }\href {\doibase 10.1007/s002200100381} {\bibfield  {journal}
  {\bibinfo  {journal} {Commun.Math.Phys.}\ }\textbf {\bibinfo {volume}
  {217}},\ \bibinfo {pages} {595} (\bibinfo {year} {2001})},\ \Eprint
  {http://arxiv.org/abs/hep-th/0002230} {arXiv:hep-th/0002230 [hep-th]}
  \BibitemShut {NoStop}%
\bibitem [{\citenamefont {Dias}\ \emph {et~al.}(2012)\citenamefont {Dias},
  \citenamefont {Horowitz},\ and\ \citenamefont {Santos}}]{Dias:2011ss}%
  \BibitemOpen
  \bibfield  {author} {\bibinfo {author} {\bibfnamefont {O.~J.}\ \bibnamefont
  {Dias}}, \bibinfo {author} {\bibfnamefont {G.~T.}\ \bibnamefont {Horowitz}},
  \ and\ \bibinfo {author} {\bibfnamefont {J.~E.}\ \bibnamefont {Santos}},\
  }\href {\doibase 10.1088/0264-9381/29/19/194002} {\bibfield  {journal}
  {\bibinfo  {journal} {Class.Quant.Grav.}\ }\textbf {\bibinfo {volume} {29}},\
  \bibinfo {pages} {194002} (\bibinfo {year} {2012})},\ \Eprint
  {http://arxiv.org/abs/1109.1825} {arXiv:1109.1825 [hep-th]} \BibitemShut
  {NoStop}%
\bibitem [{Note3()}]{Note3}%
  \BibitemOpen
  \bibinfo {note} {S.~Fischetti, D.~Marolf, and J.~Santos, `AdS flowing black
  funnels: Stationary AdS black holes with non-Killing horizons and heat
  transport in the dual CFT' {\protect \it To appear.}}\BibitemShut {Stop}%
\bibitem [{\citenamefont {Dias}\ \emph {et~al.}(2011)\citenamefont {Dias},
  \citenamefont {Horowitz},\ and\ \citenamefont {Santos}}]{Dias:2011at}%
  \BibitemOpen
  \bibfield  {author} {\bibinfo {author} {\bibfnamefont {O.~J.}\ \bibnamefont
  {Dias}}, \bibinfo {author} {\bibfnamefont {G.~T.}\ \bibnamefont {Horowitz}},
  \ and\ \bibinfo {author} {\bibfnamefont {J.~E.}\ \bibnamefont {Santos}},\
  }\href {\doibase 10.1007/JHEP07(2011)115} {\bibfield  {journal} {\bibinfo
  {journal} {JHEP}\ }\textbf {\bibinfo {volume} {1107}},\ \bibinfo {pages}
  {115} (\bibinfo {year} {2011})},\ \Eprint {http://arxiv.org/abs/1105.4167}
  {arXiv:1105.4167 [hep-th]} \BibitemShut {NoStop}%
\bibitem [{Note4()}]{Note4}%
  \BibitemOpen
  \bibinfo {note} {Instead one can also directly fix the mass $m$, imposing
  $\partial _z T |_{z=0} = m$, but we have found this numerically rather
  unstable in practice.}\BibitemShut {Stop}%
\bibitem [{\citenamefont {Bhattacharyya}\ \emph
  {et~al.}(2008{\natexlab{b}})\citenamefont {Bhattacharyya}, \citenamefont
  {Loganayagam}, \citenamefont {Mandal}, \citenamefont {Minwalla},\ and\
  \citenamefont {Sharma}}]{Bhattacharyya:2008mz}%
  \BibitemOpen
  \bibfield  {author} {\bibinfo {author} {\bibfnamefont {S.}~\bibnamefont
  {Bhattacharyya}}, \bibinfo {author} {\bibfnamefont {R.}~\bibnamefont
  {Loganayagam}}, \bibinfo {author} {\bibfnamefont {I.}~\bibnamefont {Mandal}},
  \bibinfo {author} {\bibfnamefont {S.}~\bibnamefont {Minwalla}}, \ and\
  \bibinfo {author} {\bibfnamefont {A.}~\bibnamefont {Sharma}},\ }\href
  {\doibase 10.1088/1126-6708/2008/12/116} {\bibfield  {journal} {\bibinfo
  {journal} {JHEP}\ }\textbf {\bibinfo {volume} {0812}},\ \bibinfo {pages}
  {116} (\bibinfo {year} {2008}{\natexlab{b}})},\ \Eprint
  {http://arxiv.org/abs/0809.4272} {arXiv:0809.4272 [hep-th]} \BibitemShut
  {NoStop}%
\end{thebibliography}%

\newpage

\appendix

\section{Appendix A: Illustrative toy example}

In this appendix we illustrate the old method, which assumes a Killing horizon, and the new ingoing method described in this paper, using a simple toy example; numerically finding the Schwarzschild solution assuming spherical symmetry. The purpose is to contrast the two methods, and illustrate explicitly how to use them in as simple a context as possible. We hope this will be of use to a reader interested in actually implementing these methods in more complicated settings.\\

\noindent
{\small \bf The old Killing horizon method}\\

\noindent
We write an ansatz for the black hole metric as,
\begin{equation}
\begin{aligned}
ds^2 =& - r^2 \left( \kappa^2 f B + r^2 A \right) dt^2 +  \frac{4 B d r^2}{f^4} +  \frac{S}{f^2}  d\Omega^2 
\end{aligned}
\end{equation}
where $d\Omega^2$ is the line element on the unit round 2-sphere, $A, B, S$ are smooth (at least $C^2$) functions of a compact coordinate $r \in [0, 1]$ and $f(r) = 1 - r^2$. We require $A = B = S = 1$ at $r = 1$ giving asymptotic flatness. Regularity at the horizon $r = 0$ implies that $A, B, S$ must be smooth functions in $r^2$, and then $\kappa$ gives the surface gravity.

We may discretize $A, B, S$ on the interval $[0, 1]$ and then require Neumann boundary conditions at $r = 0$ for these functions. 
As an example, one might take the reference metric and the initial guess to be the above metric with $A = B = 1 - f/2$ and $S = 1$ (which is not Schwarzschild).
On finding a solution, one obtains that odd derivatives at $r = 0$ vanish. However this approach suggests there is a boundary at the horizon, which really there is not. A better way to think is solving the problem in the domain $[-1, 1]$ requiring smooth and \emph{even} solutions so $A(-r) = A(r)$ and similarly for $B$ and $S$. Using finite difference or pseudo-spectral methods 
one may choose lattices with even numbers of points that avoid $r=0$ altogether. 
Then we have no boundary at the horizon and instead solve the problem on the complete  $t=0$ slice representing the Einstein-Rosen bridge that intersects the bifurcation surface.\\

\noindent
{\small \bf The new ingoing non-Killing horizon method}\\

\noindent
Take an ansatz
with ingoing time $t$,
\begin{eqnarray}
ds^2 =  - T dt^2 - \frac{2}{z^2} V dt dz + \frac{1}{z^4} A dz^2 + \frac{S}{z^2}  d\Omega^2
\end{eqnarray}
for a compact coordinate $z \in [0, 1]$ with $T, V, A, S$ sufficiently smooth (at least $C^2$) functions of $z$. We impose asymptotic flatness as $T = V = S = 1$ and $A = 0$ at $z=0$.
For monotonic $S$ the horizon occurs at $T = 0$ and provided $V$ is non-zero and the functions are smooth in $z$ there then the horizon will be regular. 
Now $z = 1$ is not regarded as a boundary, and the equations \eqref{eq:harmonic} are imposed there as in the interior of the domain.
We must impose one condition to select the Schwarzschild solution we wish to find, ie. to choose a mass. A simple way to fix this is that instead of solving the $vv$ component of \eqref{eq:harmonic} at $z=1$, instead we replace it with a Dirichlet condition for $T$ at $z = 1$, so $T|_{z=1} = T_{inner}$ where $T_{inner} < 0$ to ensure the domain pierces the horizon
\footnote{Instead one can also directly fix the mass $m$, imposing $\partial_z T |_{z=0} = m$, but we have found this  numerically rather unstable in practice.}.  
Consider the smooth metric,
\begin{eqnarray}
\bar{ds}^2 =  - (1 - \alpha z) dt^2 - \frac{2}{z^2} dt dz + \frac{1 + z^3/10}{z^2}  d\Omega^2
\end{eqnarray}
for constant $\alpha$ -- note this is not Schwarzschild. 
As an example let us take this metric with $\alpha = 1.10$ as the reference metric, and with $\alpha = 1.20$ as the initial guess. The Newton method then converges to a solution with $\partial_z T|_{z=0} = - 1.06$.
Note that had one tried to find a black hole where the horizon was located at $z > 1$, 
for example taking an initial guess so that $T_{inner} > 0$ then the method fails. The requirement of a smooth horizon is crucial to impose boundary conditions correctly.

Two important points arise in this example. Firstly experimentally we find that for second order finite difference the method fails, presumably as it doesn't impose smoothness of the functions to a sufficient degree. Certainly for fourth order or above the method works very well, as it does for pseudo-spectral differencing.
A second point is that if we had not imposed $T = T_{inner}$ at $z = 1$ but only the equations of motion there we would not obtain a locally unique solution. 
We emphasise that in this ingoing method the reference metric does \emph{not} determine the moduli (in this case mass) of the solutions found.

\section{Appendix B: Details for inhomogeneous plasma flows and their duals}

For a given boundary metric deformation specified by the constants $\alpha$ and $\beta$, the solution is determined by the parameters $r, z_0$ as discussed in the main text. We note that the global scaling $x^a = ( t, \rho, y ) \to \lambda \, x^a$ of the boundary together with scaling of the parameters $\alpha \to \alpha$ and $\beta \to \lambda^{-1} \beta$,  and the boundary stress tensor components, $( G_4 \langle T_{ab} \rangle) \to \lambda^{-3} (G_4 \langle T_{ab}\rangle)$ relates solutions due to the conformal invariance of the boundary theory. 
In practice we choose $z_0 = 1$ and $z_{max} = 1.025$, then vary $r$ to obtain the required ingoing velocity $v_0$ (or equivalently ingoing value of $\langle T^{t\rho}\rangle/\langle T^{tt} + T^{t\rho} \rangle$) which is scale invariant. 
In principle we would then use the above scaling to generate a solution with the required ingoing temperature $T_0$ (or equivalently ingoing value of $G_4 \langle T_{tt}\rangle$). However, for reasons that we do not understand (presumably related to the details of the way we fix the moduli of the solution) the value of $T_0$ is actually the same to better than percent level for the various values of $\beta$ we have explored, and so we have not needed to apply any scaling to the data presented here.

\begin{figure}
  \includegraphics[clip,width=3.2in]{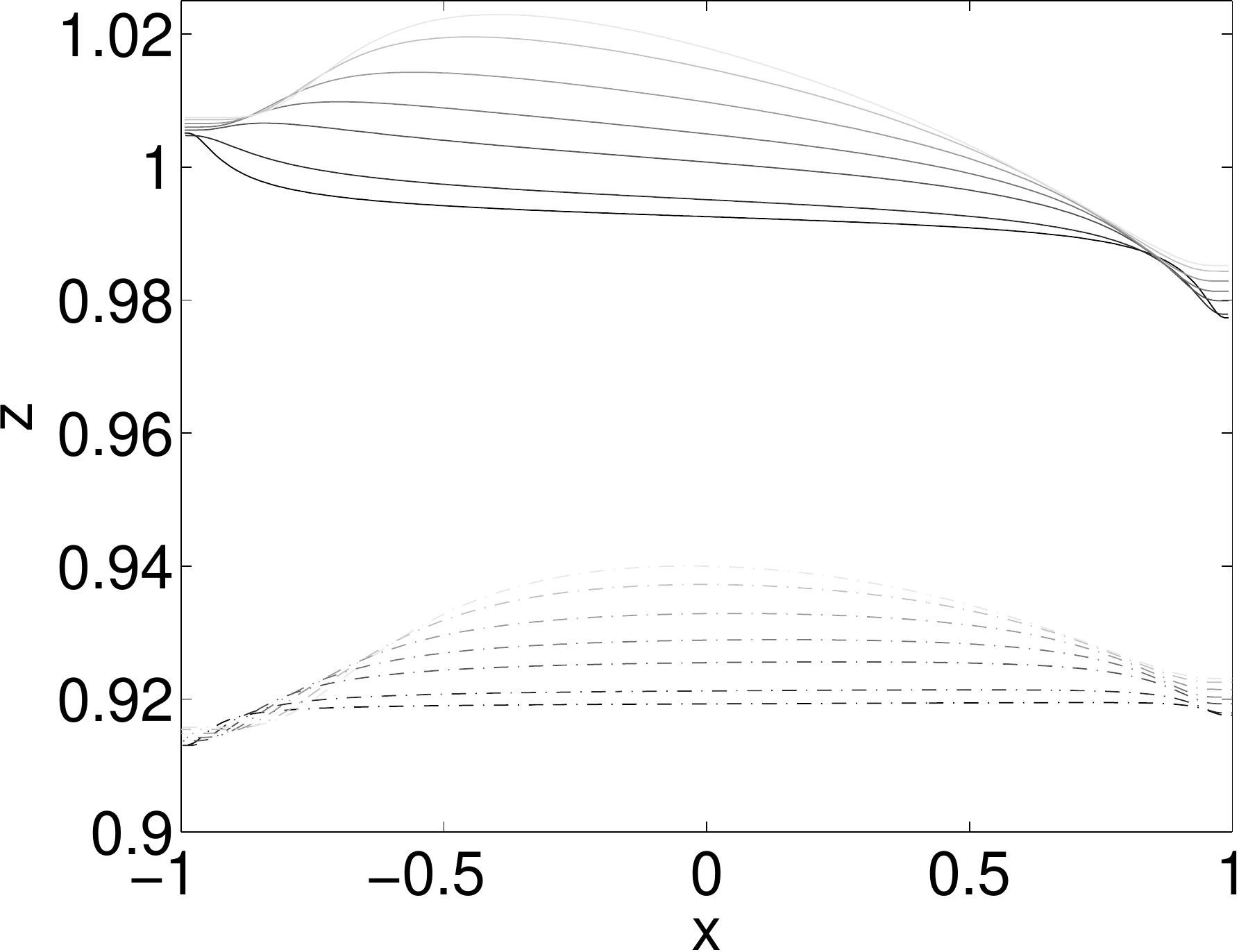}
  \caption{Coordinate positions of the horizon (solid lines) and ergosurface (dashed lines) for the flows in Fig. \ref{fig:flows}.  The larger the value of $\beta$ the deeper the horizon penetrates in the $z$-direction. Entropy is produced along the flow and hence the area density of the horizon is larger on the right end.}
\label{fig:horizon} 
\end{figure}

 In figure \ref{fig:horizon}, we plot the position of the ergoregion (defined by $T = 0$), and the horizon solved from the o.d.e. $g^{\mu\nu}\,\partial_\mu h\,\partial_\nu h = 0$ discussed in the main text. We note that both have a complicated dependence on $x$, although as expected this vanishes as $x\to \pm 1$ where the metric becomes that of a homogeneous black brane, which in the coordinates defined by our reference metric will have ergoregion and horizon at constant $z$. Note also that the position of the horizon lies entirely within our domain $0 < z < z_{max} = 1.025$ for all the solutions presented here. Since the metric functions are smooth at the position of the horizon, the future horizon is regular.

We extract the vev of the CFT stress tensor 
from the bulk solution using the standard holographic renormalisation prescription \cite{deHaro:2000xn}. We proceed by first computing the boundary stress tensor in Fefferman-Graham coordinates. Then we construct, in a near boundary expansion, the change of coordinates from Fefferman-Graham coordinates to our working coordinates \eqref{eq:metric} by requiring that $\xi^\mu=0$ order by order.  
Defining,
\begin{equation}
\frac{16\pi\,G_4}{3} \langle \left( \begin{array}{ccc} T_{tt} & T_{\rho t} & 0 \\ T_{\rho t} & T_{\rho\rho} & 0 \\ 0 & 0 & T_{yy} \end{array} \right) \rangle =\, \left( \begin{array}{ccc} t_3(\rho) & u_3(\rho) & 0 \\ u_3(\rho) & b_3(\rho) & 0 \\ 0 & 0 & s_3(\rho) \end{array} \right)\,,
\end{equation}
then $t_3, b_3, u_3, s_3$ can be obtained from the bulk solution. 
For example, for $t_3$ one finds an equation,
\begin{equation}
\label{eq:t3extract}
\begin{aligned}
&t_3(x)= \frac{1}{3}+\frac{1}{3!}\,\partial_z^3T\big|_{z=0}\\
&-\frac{f^4 \big[-4 \sigma' (-6+7 f)+f (f \sigma^{(3)} -12 x \sigma^{(2)} )\big] (101 s_r+5 s_{3r})}{576 \sigma }\\
&-\frac{f^5\,\sigma' (-f\,\sigma^{(2)}+4\,x\,\sigma' ) (387 \,s_r+47 \,s_{3r})}{1024\, \sigma ^2}\\
&-\frac{ f^6\,(\sigma')^3 (5981\, s_r+993\,s_{3r})}{32768\, \sigma ^3}\,,
\end{aligned}
\end{equation}
where $\sigma$ is defined in \eqref{eq:sigma} and we give all functions in terms of the compact coordinate $x$ (rather than $\rho$), $f(x)=1-x^2$
and $s_{kr} = \sinh{k r}$.
However, since the gauge condition $\xi^\mu = 0$ relates the derivatives of the various metric functions, there are 3 other ways to extract $t_3$ from three $z$ derivatives of the other metric functions than $T$. For a continuum solution these must all give the same answer, and we use this to test the accuracy of our stress tensor determination shortly.
The other $b_3, u_3, s_3$ can similarly be extracted from three $z$ derivatives of the various metric functions, and again there may be multiple ways to do this which are equivalent on Einstein solutions. 
The equations of motion also imply that $b_3, u_3, s_3$ are locally related to $t_3$ as the stress tensor is traceless and conserved, which again we check shortly.

\begin{figure}
\includegraphics[clip,width=3.5in]{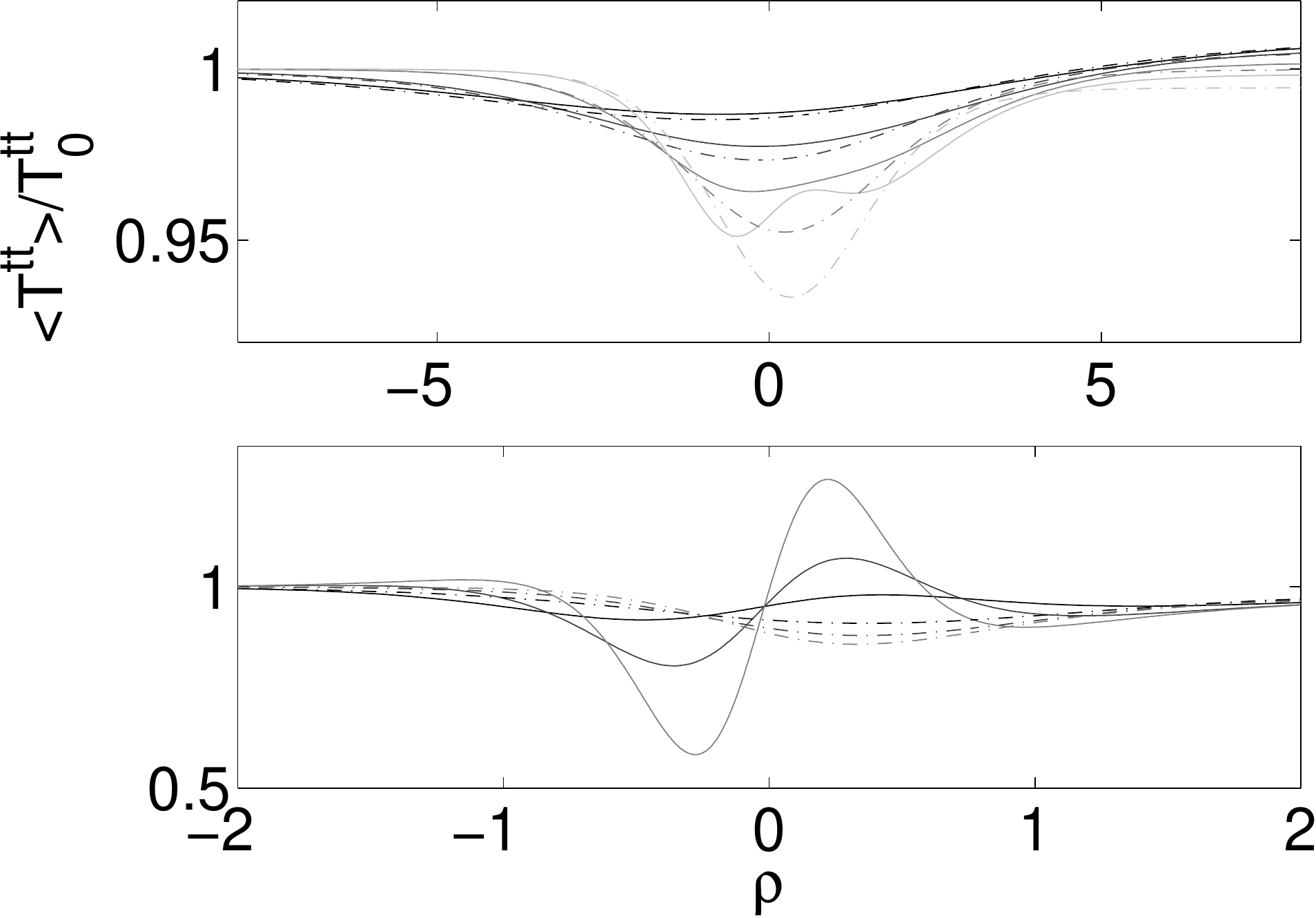}
\caption{Plot of $\langle T^{tt}\rangle$ normalised by its ingoing value $T^{tt}_0$ vs. $\rho$ for the holographic stress tensor (solid lines) and the stress tensor of viscous hydrodynamics (dashed lines). \textit{Top}:  Flows as in figure \ref{fig:flows} with $\beta=0.2,0.3,0.5,0.7$. \textit{Bottom}: flows with $\beta=1.,1.5,2$. For $\beta\simeq O(1)$ the stress tensor exhibits $O(1)$ features at small scales (compared to the length scale set by the temperature) and hydrodynamics no longer provides a valid description of the flow.}
\label{fig:Ttt}
\end{figure}

In figure \ref{fig:Ttt} we display the vev of the 
$T^{tt}$ component of the holographic stress tensor and compare it with the same component of the viscous hydrodynamic stress tensor,  \eqref{eqn:hydrostresstensor}, for varying $\beta$ in analogy with  figure \ref{fig:flows} in the main text. We see the same agreement with viscous hydrodynamics at small $\beta$, with strong deviations from it for $\beta \sim O(1)$.
We note that $\langle T^{tt} \rangle$ is well behaved even for the $\beta = 2$ flow which has a superluminal region.

\begin{figure}
\includegraphics[clip,width=3.5in]{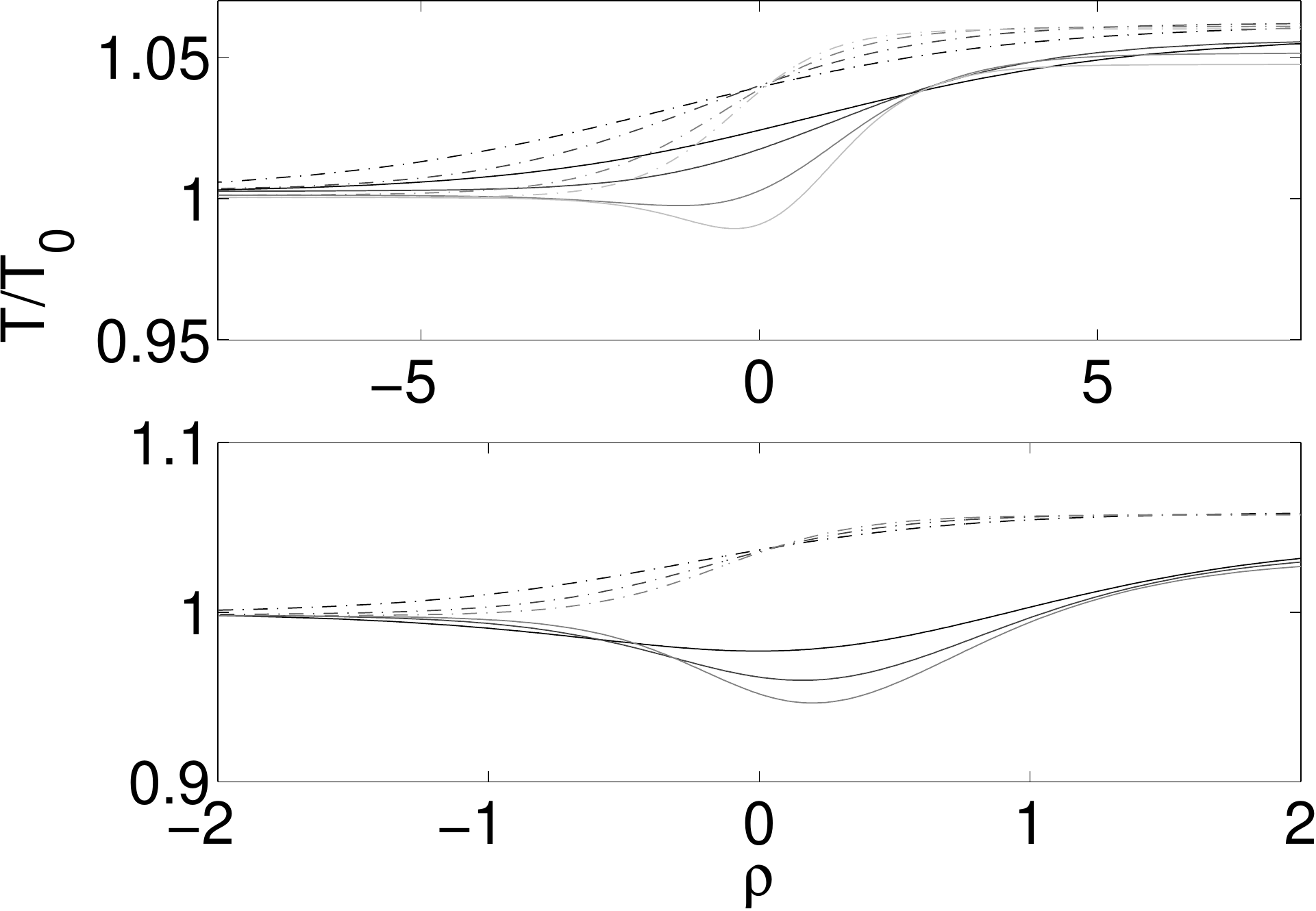}
\caption{
Temperature, $T$, plotted normalised by the ingoing temperature $T_0$ vs. $\rho$ from viscous fluid/gravity hydrodynamics (solid lines) and for  ideal fluid/gravity hydrodynamics (dashed lines) used to compare to the holographic plasma flows in figures \ref{fig:flows} and \ref{fig:Ttt}. 
We emphasise that these only approximate the holographic plasma flow well for small $\beta$.
\textit{Top:} flows with $\beta=0.2,0.3,0.5,0.7$. \textit{Bottom:} flows with $\beta=1.,1.5.,2$. 
We see for small $\beta$ agreement between the viscous and ideal hydrodynamics since there is little entropy generation. For larger $\beta \sim O(1)$ viscosity becomes important and the behaviours strongly differ.
}
\label{fig:temp}
\end{figure}

In the figures above we have compared the holographic plasma behaviour extracted from the dual black holes to the viscous hydrodynamics predicted by the fluid/gravity correspondence. This gives a good approximation for $| \beta / T_0 | \ll 1$. We now give details about these hydrodynamic fluid flows.
Recall that the viscous fluid approximation to the plasma flow from fluid/gravity is determined by the fluid stress tensor\cite{Bhattacharyya:2008mz},
\begin{eqnarray}
16 \pi && G_4  \langle T^{ab}\rangle =  \left( \frac{4}{3} \pi T \right)^3  \left( u^a u^b + \frac{1}{2} P^{ab} \right)  \\
 && - 2 \left( \frac{4}{3} \pi T \right)^2 P^{ac} P^{bd} \left( \nabla_{(c} u_{d)} - \frac{1}{2} g^{(b)}_{cd} \nabla^e u_{e} \right) + O\left(\nabla^2 u \right) \nonumber
\label{eqn:hydrostresstensor}
\end{eqnarray}
for a 1+2-dimensional boundary metric $ds^2 = g^{(b)}_{ab} dx^a dx^b$,  
temperature $T$, 3-velocity $u^a$ with $u^2 = -1$ and with
$P_{ab} =  u_a u_b + g^{(b)}_{ab}$.
The first term is that of an ideal fluid, and the latter is due to shear viscosity.
For our flows, we take the 3-velocity in the $\rho$ direction, so $u^a = ( \gamma , \gamma v , 0 )$, with $v$ the velocity and $\gamma^{-2} = 1-v^2$.
For small but non-vanishing $\beta$ the viscous term will generate entropy and the fluid deviates from ideal behaviour. 
The equation of motion for the fluid is given by conservation of this stress tensor. 
For our boundary metric there are two non-trivial components. One immediately yields $G_4 \langle T_{tx} \rangle \propto 1/\sqrt{\sigma}$. Combined with the second, one obtains first order o.d.e.s for the fluid velocity and temperature, as discussed in \cite{Khlebnikov:2010yt,Khlebnikov:2011ka}. It is these we have solved in order to compare to our numerical bulk solutions.
In figure \ref{fig:temp} we display the behaviour of the temperature for the flows obtained from viscous hydrodynamics which we compared to the stress tensor from the gravity dual in figures \ref{fig:flows} and \ref{fig:Ttt} (we note that the velocity of these hydrodynamic flows are already shown in figure \ref{fig:flows}).
For comparison with the viscous hydrodynamics we also show the ideal hydrodynamics solutions for the same ingoing fluid data. We see that as expected, for small $\beta$ these closely agree, but for $\beta \sim O(1)$ the viscous behaviour departs strongly from the ideal behaviour, and likewise as we have seen in figures \ref{fig:flows} and \ref{fig:Ttt}, deviates from the full plasma behaviour as deduced from the gravity.

\section{Appendix C: Numerical errors and metric functions}

For the results presented in this paper we have discretized the harmonic Einstein equation using sixth order finite differencing, taking a uniform grid with $N_z$ lattice points in the $z$ direction, and $N_x = 4 N_z$ lattice points in $x$. We note that especially for the small $\beta$ solutions where there are sharp gradients in the function $\sigma$ near the boundaries of the domain it is important to have sufficient resolution in the $x$ direction to obtain accurate results. We have used resolutions up to $N_z \times N_x = 70 \times 280$ points. Typically we begin by finding solutions at lower resolutions, and then use these as initial data to find the higher resolution solutions.

\begin{figure}
  \includegraphics[clip,width=3.in]{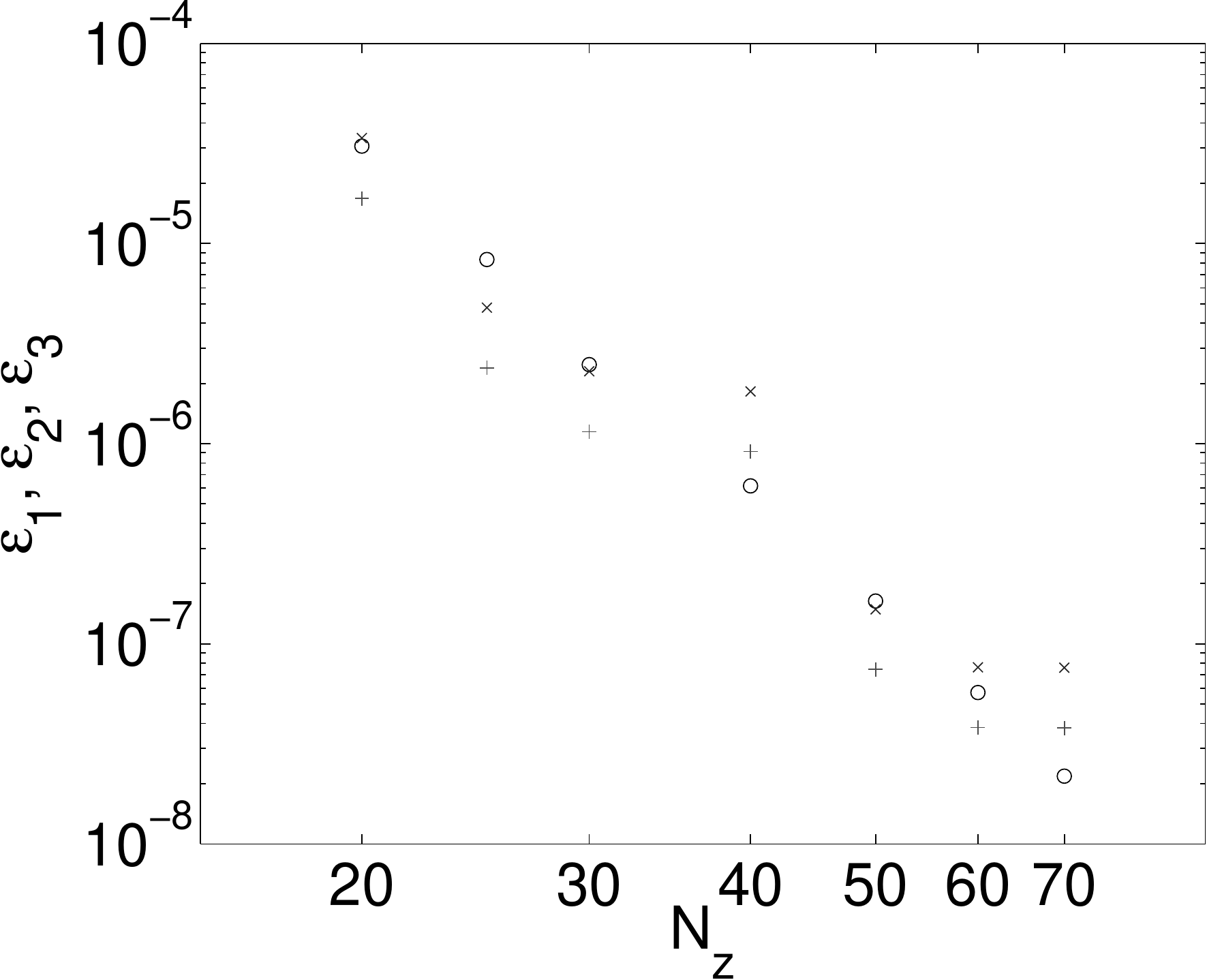}
  \caption{
  Convergence plots for the solution with $\beta=1$. We show $\mathcal E_1$ (`o'), $\mathcal E_2$ (`x') and $\mathcal E_3$ (`+'), which measure the maximum error in the Ricci scalar, Ricci tensor and the magnitude of $\sqrt{ | \xi^\mu \xi_\mu |}$ respectively, 
   as a function of the number of grid points in the $z$-direction, $N_z$. The resolution in $x$ is given as $N_x = 4 N_z$. We see linear convergence in this log-log plot. For $\mathcal E_1$ and $\mathcal E_2$ we find the slope is $\sim 4$ indicating fourth order convergence, whilst for $\mathcal E_3$ the slope is $\sim 6$. Other solutions exhibit the same convergence behaviour.}
\label{fig:convergence}
\end{figure}

We characterise the numerical error in our solutions by computing the error in solving the Einstein equations as,
\begin{equation}
\begin{aligned}
\mathcal{E}_1 &= \max_{0 < z < H(x)}  \left| \frac{R}{12} + 1 \right| \,,  \\
\mathcal{E}_2 &= \max_{0 < z < H(x)}  \left| \frac{R_{\mu\nu} R^{\mu\nu}}{36} - 1 \right| \,, \\
\mathcal{E}_3 &= \max_{0 < z < H(x)}  \sqrt{ \left|  \xi^\mu \xi_\mu \right| } 
\end{aligned}
\end{equation}
where each is a scalar quantity which should vanish in the continuum for a solution, and is maximised for the solution in the exterior of the horizon. We maximise only over the 
region exterior to the horizon
 to obtain a well defined geometric quantity. We note that similar results are obtained
when maximising over the entire domain. 
Typical results of convergence tests are shown for an intermediate value of $\beta = 1$ in figure \ref{fig:convergence}. 
For the maximum resolutions used, we see that the maximum local error in the solution is better than $10^{-7}$, as stated in the text. 
In addition, $\mathcal E_3 \to 0$ in the continuum limit, which indicates that our solutions are not Ricci solitons. Similar results are obtained for the other values of $\beta$ (including $\beta = 2$) discussed in this paper.

We note that whilst we have used sixth order finite differencing, the slope of these curves against $\log N_z$ is between $\sim 4 - 6$ depending on the quantity. We would naively expect $\sim 6$ for smooth solutions. We believe the observed lack of smoothness is not physical but due to our coordinate choice near the boundaries $x = \pm 1$. Since these boundaries are regular singular points of the p.d.e.s it maybe that there are $(x \mp 1)^p \log| x \mp 1 |$ behaviours in the expansions of the metric functions for our gauge choice, where $p$ is some power presumably with $p \ge 4$. We emphasise that we require only second derivatives to be defined for a solution to the Einstein equations, and the convergence we see is certainly much better than that, indicating the metric functions are better than $C^2$ in smoothness. As we show later, explicit calculation of the various two derivatives of metric functions gives well behaved results, again confirming better than $C^2$ smoothness. However, the apparent lack of $C^\infty$ smoothness does lead to poor convergence results when using pseudo-spectral differencing, hence our use of finite difference. An obvious future direction is to improve the coordinate choice. 

\begin{figure}
  \includegraphics[clip,width=3.in]{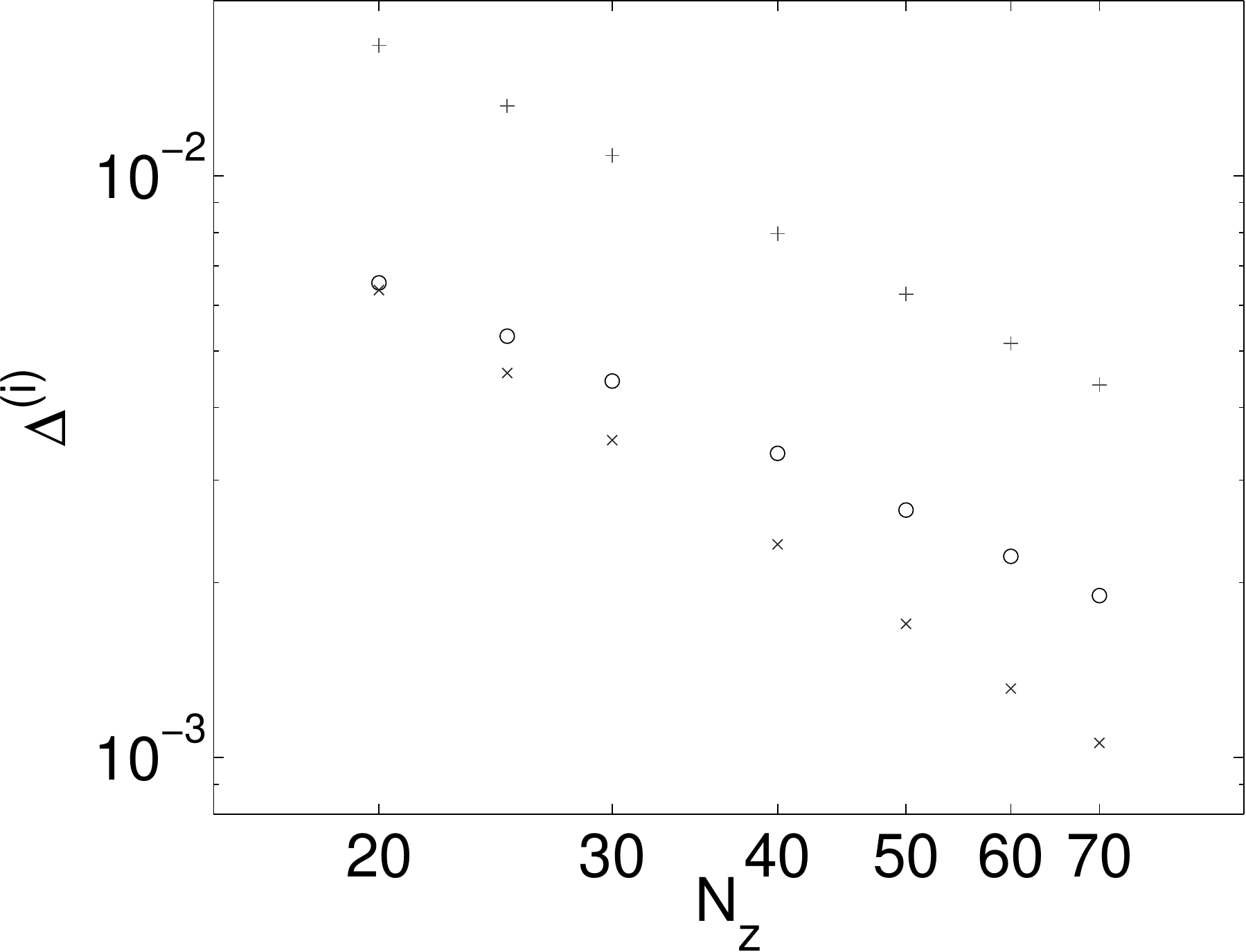}
\caption{ The function $t_3$ may be extracted from the metric functions in 4 independent ways which are equivalent in the continuum. Here we plot the maximum value of the three fractional errors, ${\Delta}^{(i)}$ characterising the deviation between these
as a function of the number of grid points in the $z$ direction, $N_z$, in a log-log scale for the $\beta = 1$ solution. We observe first order (or slightly better) convergence, which is consistent with the overall observed $4^{\textrm{th}}$ order convergence as the quantity $t_3$ involves 3 derivatives of the metric functions. We obtain the same results for other values of $\beta$.}
\label{fig:fracerror}
\end{figure}

We monitor the errors in the extraction of the stress tensor. 
As discussed in appendix B the components of the stress tensor $t_3(x)$, $u_3(x)$, $s_3(x)$ and $b_3(x)$ may be extracted from the 7 metric functions in different ways which should agree in the continuum.
In figure \ref{fig:fracerror} we display the fractional error in the different ways of extracting $t_3$:
\begin{equation}
\Delta^{(i)}=\max_{-1<x<1}\left|1-\frac{t_3^{(i)}}{t_3^{(0)}}\right|\,,\quad i=1,2,3\,,
\end{equation}
where $t_3^{(0)}$ denotes the value of $t_3$ obtained from $\partial_z^3T|_{z=0}$ as given in equation \eqref{eq:t3extract},
and $t_3^{(i)}$ correspond to the other values obtained from the remaining independent combinations of metric functions.  
As this figure shows, $\Delta^{(i)}$ is consistent with vanishing in the continuum limit with a slope  $\sim 1$ in a log-log plot. This is the expected behaviour;  from the equations of motion we have seen that we have fourth order convergence and the calculation of $t_3$ involves taking three derivatives of the metric functions. Therefore we expect the error in this quantity should exhibit approximately first order convergence.  We see for our highest resolution data that the maximum fractional error is less than percent level. We obtain analogous results for other components of the stress tensor which may be extracted in multiple ways, which we note includes the test of tracelessness of the stress tensor.

\begin{figure}
  \includegraphics[clip,width=3.5in]{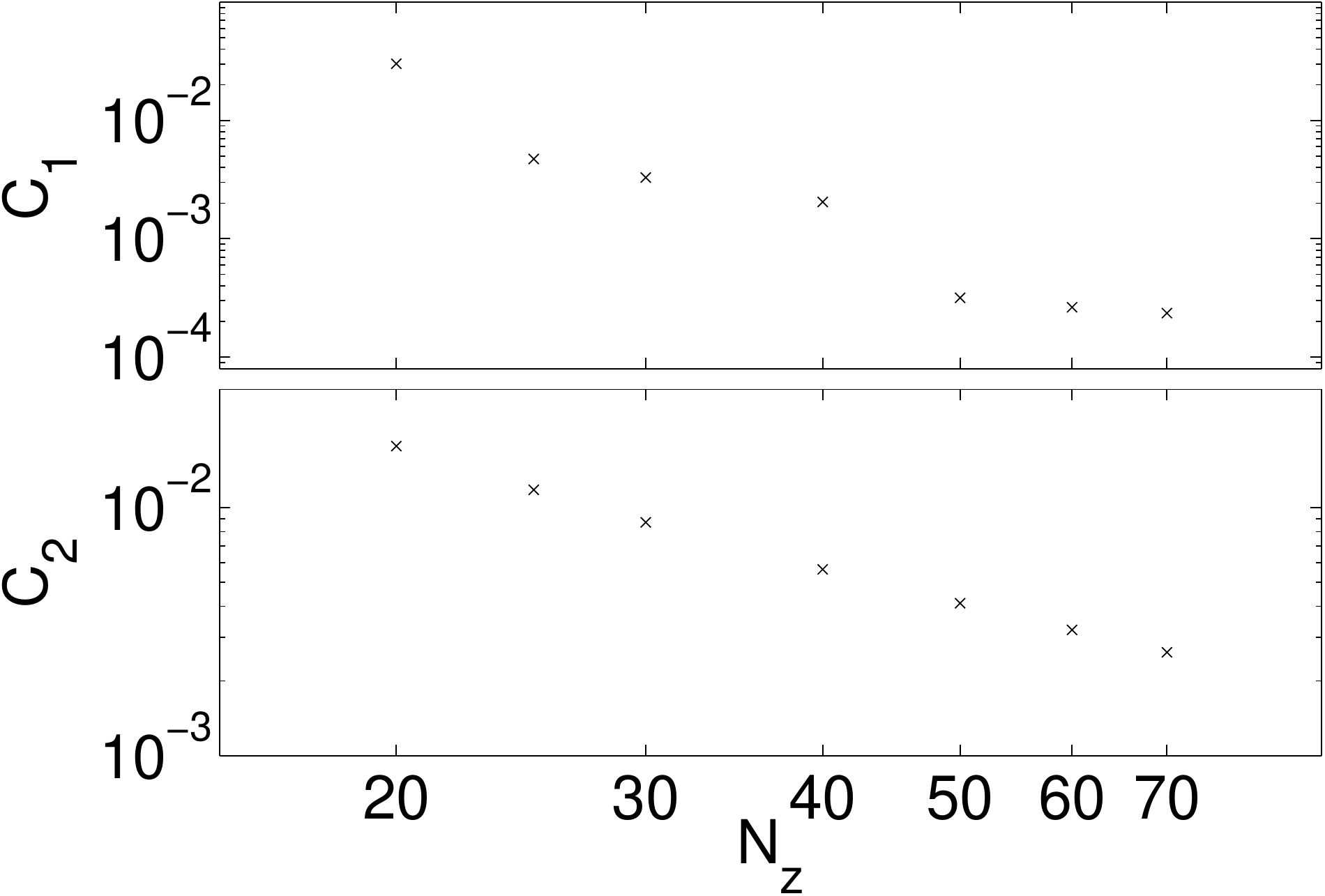}
\caption{
Convergence plots for  $\mathcal{C}_{1,2}$, the maximum errors in the two non-trivial components of the conservation equation for the stress tensor for the $\beta = 1$ solution. The apparent linear convergence in this log-log plot against $N_z$ has slope $\sim 4$ for $\mathcal C_1$, and  $\sim 1$ for $C_2$. These are consistent (or better) than expected, given the overall $4^{\textrm{th}}$ order convergence, and that the conservation requires three derivatives of the metric functions in $z$ and one in $x$.}
\label{fig:conservation}
\end{figure}

Next we consider the error in the two non-trivial components of the  conservation equation of the stress tensor:
\begin{equation}
\begin{aligned}
\mathcal C_1&=\max_{-1<x<1}\left|\frac{u_3'}{u_3}+\frac{1}{2}\,\frac{\sigma'}{\sigma}\right|\,,\\
\mathcal C_2&=\max_{-1<x<1}\left|\frac{b_3'}{b_3}+\frac{\sigma'}{\sigma}\left(1+\frac{t_3}{2\,b_3}\right)\right|\,,
\end{aligned}
\label{eqn:conservation}
\end{equation}
where each quantity should vanish in the continuum limit. In figure \ref{fig:conservation} we display $\mathcal C_1$ and $\mathcal C_2$ as a function of the number of grid points in the $z$ direction, $N_z$, again in a log-log plot. As this figure shows, the error in $\mathcal C_1$ exhibits almost fourth order convergence, which is better that one might have naively expected. 
On the other hand, the convergence in  $\mathcal C_2$ is slightly better than first order, which is consistent with behaviour of the error in extracting $t_3$ and $b_3$ as discussed above. We obtain similar convergence results for the other $\beta$ studied, including $\beta = 2$.

In summary, our analysis of the numerical errors show that the solutions we present are of high quality, the maximum fractional error in the solutions being better than $\sim 10^{-7}$. The finite differencing method we have implemented gives convergence to the continuum limit consistent with fourth order scaling. Our extraction of the components of the stress tensor exhibits the expected convergence, and we may estimate that the maximum error in these components is better than $1\%$.

\begin{figure}
\includegraphics[clip,width=3.5in]{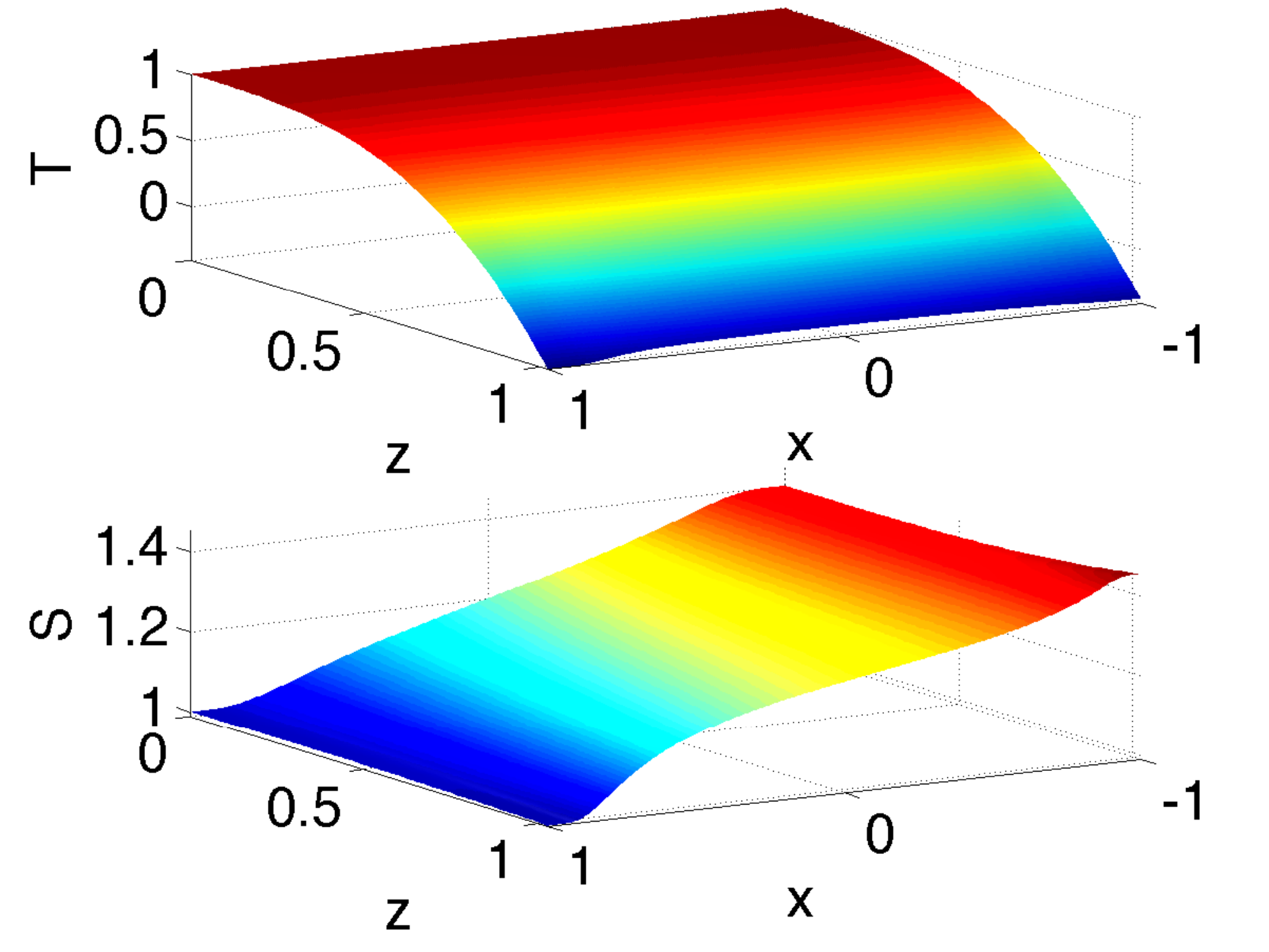}
\caption{
Metric functions $T(z,x)$ and $S(z,x)$ for the solution with $\beta =1$. These and the other remaining metric functions $V, B, F, U, A$, are well behaved everywhere in our domain, including the region inside the horizon. }
\label{fig:TS}
\end{figure}

\begin{figure}
\includegraphics[clip,width=3.5in]{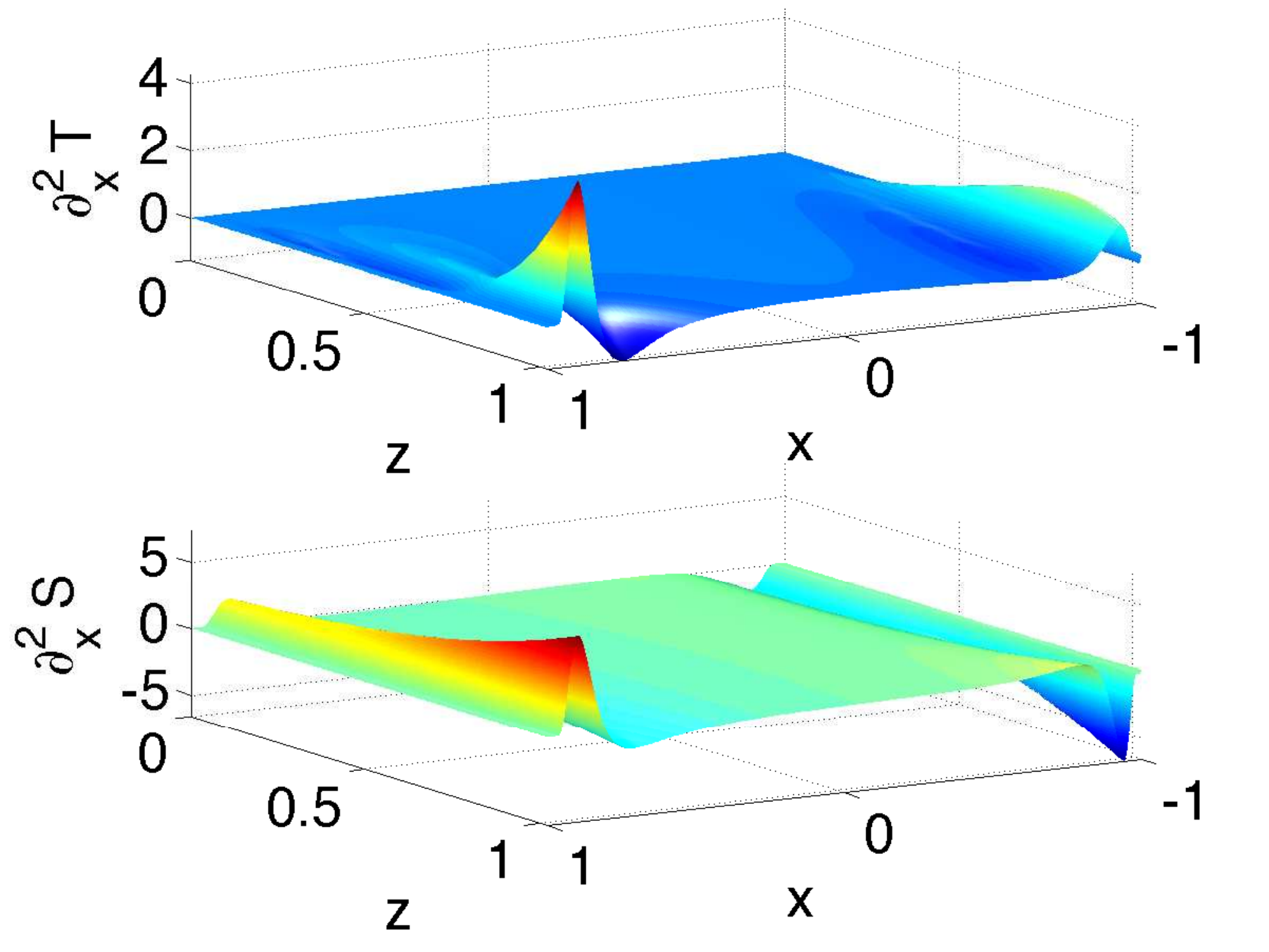}
\caption{
$\partial_x^2T$ and $\partial_x^2S$ for the $\beta =1$ solution. As expected these are largest where $\sigma(x)$ changes most rapidly.  We emphasise that these and other two derivatives of the various metric functions are well behaved over the domain and vanish at the asymptotic ends $x\to \pm 1$ where the flows and dual black brane become homogeneous. }
\label{fig:TxxSxx}
\end{figure}

We now turn to the metric functions. For concreteness, in figure \ref{fig:TS} we show $T$ and $S$ over the domain for the $\beta = 1$ solution, and note that these are coordinate scalars with respect to $z$ and $x$ coordinate transformations. 
In figure  \ref{fig:TxxSxx} we show the functions $\partial^2_x T$ and $\partial^2_x S$ to illustrate that the metric is better than $C^2$ and also that derivatives of the metric functions vanish as expected at $x \to \pm 1$ in the coordinate system defined by our reference metric (since the bulk solution becomes a homogeneous black brane there). The other metric functions show the same behaviour as those we show here. Likewise, taking other two derivative combinations of these we obtain analogously well behaved functions. We obtain similarly well behaved metric functions for all other values of $\beta$ presented here, including $\beta = 2$.

\begin{figure}
\includegraphics[clip,width=3.5in]{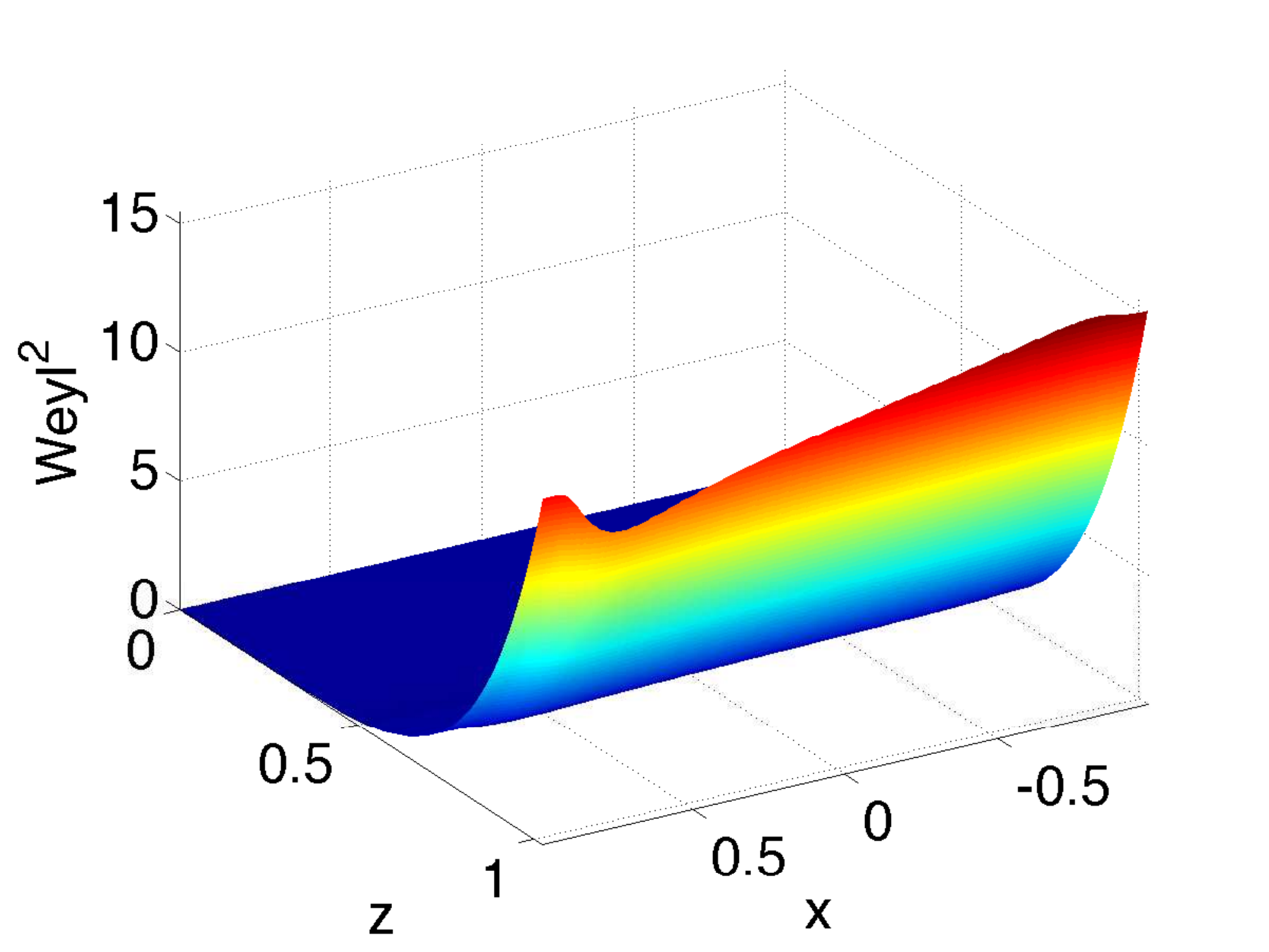}
\caption{$C_{\mu\nu\rho\sigma} C^{\mu\nu\rho\sigma}$ for the $\beta =1$ solution. This function vanishes at the boundary of AdS and well behaved elsewhere, indicating the absence of singularities. At $x\to \pm 1$ where the flow and dual black brane are homogeneous the Weyl tensor also becomes homogeneous, with its $x$-derivative vanishing.  }
\label{fig:Weyl}
\end{figure}

Finally we plot the Weyl curvature as characterised by the scalar $C_{\mu\nu\rho\sigma} C^{\mu\nu\rho\sigma}$ over our domain in figure \ref{fig:Weyl} and note that it is smooth, with no indication of any singular behaviour over the domain, again indicating the metric functions are better than $C^2$ smooth.

\end{document}